\documentclass{aastex63}

\shorttitle{Phase resolved analyses of mHz QPOs in 4U 1636-53 using HHT}
\shortauthors{Hung-En Hsieh and Yi Chou}

\graphicspath{{./}{figures/}}
\begin{document}

\title{Phase-resolved analyses of mHz quasi-periodic oscillations in 4U 1636-53 using Hilbert-Huang transform}

\author{Hung-En Hsieh}
\author{Yi Chou}
\affiliation{Graduate Institute of Astronomy, National Central University, Jhongli 32001, Taiwan}
\email{m1029003@astro.ncu.edu.tw, yichou@astro.ncu.edu.tw}



\begin{abstract}

We present phase-resolved spectroscopies based on Hilbert-Huang transform (HHT) for millihertz quasi-periodic oscillations (mHz QPOs) in 4U 1636-536. This $\sim$8 mHz QPO can be detected approximately several thousand seconds before a type-I X-ray burst. It was interpreted as marginally stable burning on the neutron-star surface. In this study, we used HHT to analyze the data collected by \emph{XMM-Newton} between 2007 and 2009. HHT is a powerful tool that enables us to obtain instantaneous frequency, amplitude and phase of non-stationary periodicity phenomena, such as QPOs. With well-defined phases, the oscillation profile of the $\sim$8 mHz QPO for 4U 1636-53 can be precisely revealed. In addition to the oscillation profile, phase-resolved spectra for the complete cycle are constructed. From the correlation between spectral parameters and fluxes, we find that the oscillation is mainly attributed to the variations of emitting area of blackbody radiation in three out of four observations with mHz QPO detections whereas the other one shows concurrent variation of temperature and flux with a constant emitting area. Although the cause of the difference is not clear, it might be related to the spectral state of the source that can be observed from hard color difference in color-color diagram.

%origin abstract
%This example manuscript is intended to serve as a tutorial and template for
%authors to use when writing their own AAS Journal articles. The manuscript
%includes a history of \aastex\ and documents the new features in the
%previous versions as well as the new features in version 6.3. This
%manuscript includes many figure and table examples to illustrate these new
%features.  Information on features not explicitly mentioned in the article
%can be viewed in the manuscript comments or more extensive online
%documentation. Authors are welcome replace the text, tables, figures, and
%bibliography with their own and submit the resulting manuscript to the AAS
%Journals peer review system.  The first lesson in the tutorial is to remind
%authors that the AAS Journals, the Astrophysical Journal (ApJ), the
%Astrophysical Journal Letters (ApJL), and Astronomical Journal (AJ), all
%have a 250 word limit for the abstract\footnote{Note that manuscripts 
%submitted to the new Research Notes of the American Astronomical Society 
%(RNAAS) do \textbf{not} have abstracts.}.  If you exceed this length the
%Editorial office will ask you to shorten it. This abstract has 180 words.

\end{abstract}



\section{Introduction} \label{sec:intro}

Various kinds of quasi-period oscillations (QPOs) that exhibit boxer peaks in the power spectra have been observed in many low mass X-ray binaries (LMXBs). These QPOs are believed to reflect phenomena associated with accretion disks (see \citealt{2006csxs.book...39V} for extensive review). However, a new kind of QPO, referred to as mHz QPO, was discovered in the accreting neutron star binaries 4U 1636-536 and 4U 1608-52 and was occasionally observed in Aql X-1 \citep{2001A&A...372..138R}. \citet{2001A&A...372..138R} suggested that these mHz QPOs are caused by a special mode of nuclear burning on the neutron star surface because the flux in 4U 1608-52 was in a narrow range during occurring mHz QPO, that is approximately consistent with the flux in which type-I X-ray bursts cease to exist. However, they did not exclude the possibility that the mHz QPOs were attributed to the instability of the accretion disk. \citet{2001A&A...372..138R} also discovered that all mHz QPOs occur before Type-I burst and disappear afterward. Following this discovery, the mHz QPOs were also detected in numerous LMXBs, such as 4U 1323-62 \citep{2012AAS...21924903S}, IGR J17480-2446 in Terzan 5 \citep{Linares_2012} and "Clocked" Burster GS 1826-238 \citep{2018ApJ...865...63S}. \citet{Yu_2002} suggested that the kHz QPO frequency in 4U 1608-52 is anticorrelated with the count rate during a mHz QPOs cycle. Because the kHz QPO is related to the Keplerian orbital frequency at the inner edge of the disk, if mHz QPO was produced from the modulation of the disk, the positive correlation between mHz QPO and kHz QPO should have been detected. This finding provides evidence that mHz QPOs comprise a phenomenon that occurs on the neutron star surface instead of the accretion disk. \citet{2008ApJ...673L..35A} found that the QPO frequency decreases with time during a mHz oscillation until it disappears as Type-I X-ray burst occurs in 4U 1636-53.

Previous studies showed that mHz QPOs associate with nuclear burning on a neutron star surface \citep{2001A&A...372..138R,2008ApJ...673L..35A}.\ The fuel of nuclear reaction is obtained from accretion.\ Therefore, the accretion rate is an important factor for nuclear burning on the neutron star surface.\ The stable burning indicates that the nuclear burning rate ($\varepsilon_{nuc}$) and cooling rate ($\varepsilon_{cool}$) are balanced, and both depend on temperature and the depth of the fuel layer \citep{1981ApJ...247..267F,10.1093/mnrasl/slv167}. The fuel piled up on the layer is compressed by a strong gravitational force, and then increase temperature and density in the fuel layer. When the ignition condition ($\varepsilon_{nuc}>\varepsilon_{cool}$) is reached, the thermonuclear runaways would occur. The result of thermonuclear runaways makes the luminosity increase in a short time. This phenomenon is called Type-I X-ray burst. Most Type-I X-ray bursts sources have mixed fuel (hydrogen and helium) on the neutron star surface, and the helium burning is ignited by the hydrogen burning. For a high accretion rate ($>11\%\dot M_{Edd}$), the accretion provides sufficient fuel to the burning layer. The hydrogen burning dominates in the fuel layer, and helium accumulates on the bottom of the layer. As the helium burning is ignited, the 3$\alpha$ reaction releases more energy and creates more carbon, that enhances the hot CNO cycle.\ The hydrogen burning also promotes 3$\alpha$ reaction.\ This burning layer increases $\varepsilon_{nuc}$ further, and then runaway thermonuclear reactions occur while reaching the ignition condition \citep{2017arXiv171206227G}.

\citet{2007ApJ...665.1311H} proposed that the mHz QPO occurs in a special condition between stable and unstable burning, that is referred to as marginal burning. The nuclear reaction rate increases as the temperature in the fuel layer increases, that causes the fuel in the layer being consumed faster than being supplied by the accretion. Thus, the thickness of the burning layer, $y$, decreases. The cooling rate is proportional to $F/y$, where $F$ is the outward flux. The outward flux $F\approx acT^{4}/3\kappa y$, where $a$ is the radiation constant, $c$ is the speed of light and $\kappa$ is the opacity \citep{1998ASIC..515..419B}. Therefore, as the thickness of the burning layer $y$ decreases, the cooling rate increases to reduce the temperature and the nuclear reaction rate. As the rate of fuel being consumed by the nuclear reaction is less than being supplied by the accretion, the thickness of the burning layer $y$ increases, that causes a smaller cooling rate. Therefore, the temperature and nuclear reaction rate increase again. \citet{2007ApJ...665.1311H} estimated that it takes about $P \cong \sqrt{t_{therm}t_{acc}} \sim120$ sec to complete this QPO cycle.  According to this model, marginally stable nuclear burning occurs at accretion rate close to the Eddington rate \citep{1998ASIC..515..419B, 2007ApJ...665.1311H}. However, the observation results showed that the accretion rates are much lower than the Eddington rate in 4U 1608-52 and 4U 1636-536 \citep{2001A&A...372..138R}. \citet{2009A&A...502..871K} attempted to find a possible solution by mixing processes. Furthermore, the width of the interval of the accretion rates between stable and unstable nuclear burning strongly depends on the composition of the layer and reaction rates \citep{2014ApJ...787..101K}.

\citet{2014MNRAS.445.3659L} and \citet{2015MNRAS.454..541L} analyzed the mHz QPOs in 4U1636-536 using \emph{XMM-Newton} observations. They also observed the drift in the QPO frequency, that is positively correlated with the temperature of the blackbody component from the neutron star surface. Such drifts could be caused by the cooling of the burning layers. By analyzing 39 Type-I X-ray bursts and mHz QPO in 4U 1636-536, \citet{2016MNRAS.463.2358L} determined that the mHz QPO can only be detected when the convexity of the corresponding Type-I X-ray burst is positive. According to the model proposed by \citet{2008MNRAS.383..387M} and \citet{Cooper_2007}, the convexity links to the ignition site of the bursts on the neutron-star surface. \citet{2016MNRAS.463.2358L}  suggested that the marginally stable nuclear burning process may occur at the neutron-star equator. Through phase-resolved spectroscopy of mHz oscillations of 4U 1636-536, \citet{2016ApJ...831...34S} concluded that the mHz QPOs are caused by a variable surface area of nuclear burning on neutron star surface with constant temperature. Nevertheless, \citet{2018ApJ...865...63S} detected the oscillation of the blackbody temperature component in an effectively constant emitting area in 8 mHz QPO of GS 1826-238.

Phase-resolved analysis may help us to further understand the nature of mHz QPOs. However, the definition of the phase with traditional analysis methods (e.g.,\ Fourier transform) for the unstable oscillation phenomena, such as QPOs, is difficult. In this research, we attempted to use Hilbert-Huang transform (HHT) to analyze mHz QPOs. The HHT is a powerful tool for analyzing the non-stationary periodicity phenomenon and has been successfully applied in astronomical research, such as the QPO in the active galactic nucleus RE J1034+396 \citep{2014ApJ...788...31H} and the 4 Hz QPO in the black hole X-ray binary XTE J1550-564 \citep{2015ApJ...815...74S}. The HHT enables us to not only trace the variation of frequency for the QPO but also process phase-resolved analyses even though the periodicity is unstable. In this paper, we present our analysis of the evolution of the frequency of mHz QPOs in 4U1636-53 and the QPO phase-resolved spectral variations. In Section \ref{sec:obs}, we briefly introduce the \emph{XMM-Newton} observations and data reduction process. In Section \ref{sec:analysis}, we describe how HHT analysis is applied in mHz QPOs and demonstrate the phase-resolved spectral analysis that reveals the spectral parameter variations for the complete cycle. From the correlation between spectral parameters and fluxes, we find that the oscillation is mainly attributed to the variations of emitting area of blackbody radiation in three out of four observations with mHz QPO detections whereas the other one shows concurrent variation of temperature and flux with a constant emitting area. Finally, we will discuss the HHT-based timing property and possible implications about our spectral analysis in Section \ref{sec:discussion}.

%\section{Manuscript styles} \label{sec:style}
\section{Observation} \label{sec:obs}

The data being analyzed with the HHT in the work were collected by \emph{XMM-Newton} between 2000 and 2015, a total of 12 observations as listed in Table 1. The HHT favors the data detected by high-altitude X-ray telescopes, such as \emph{XMM-Newton} because the observation gaps from low-altitude satellites would induce too much alias. For 4U 1636-53, \emph{XMM-Newton} can provide $\sim$30,000s continuous exposures, that are sufficient for the HHT analysis of $\sim$8 mHz QPOs. The data analyzed in this work were collected by the European Photon Imaging Camera (EPIC-PN) of \emph{XMM-Newton}, and the data reduction was performed by the Science Analysis System (SAS) version 16.0.0. The Current Calibration Files (CCF) of XMM-Newton in this work is updated to XMM-CCF-REL-371. Similar to the selection criteria proposed by \citet{2015MNRAS.454..541L}, we selected a 10 columns wide region that is centered at the position of the source and only single and double events (PATTERN $\leq$ 4). The xmmselect, the graphical interface to the SAS {\tt extractor}, produced event files comprising photon energies 0.8 to 11 keV with a time resolution of 0.03 milliseconds in the timing mode.

%%%%% HHT %%%%%%
\section{Data Analysis and Results} \label{sec:analysis}
\subsection{Hilbert-Huang Transform Analysis} \label{subsec:HHt}

\begin{figure*}[ht!]
%\includegraphics[width=0.4\linewidth]{dynamic2007.pdf}
%\includegraphics[width=0.4\linewidth]{dynamic2008.pdf}
%\\
%\includegraphics[width=0.4\linewidth]{dynamic200903.pdf}
%\includegraphics[width=0.4\linewidth]{dynamic200909.pdf}
\gridline{\fig{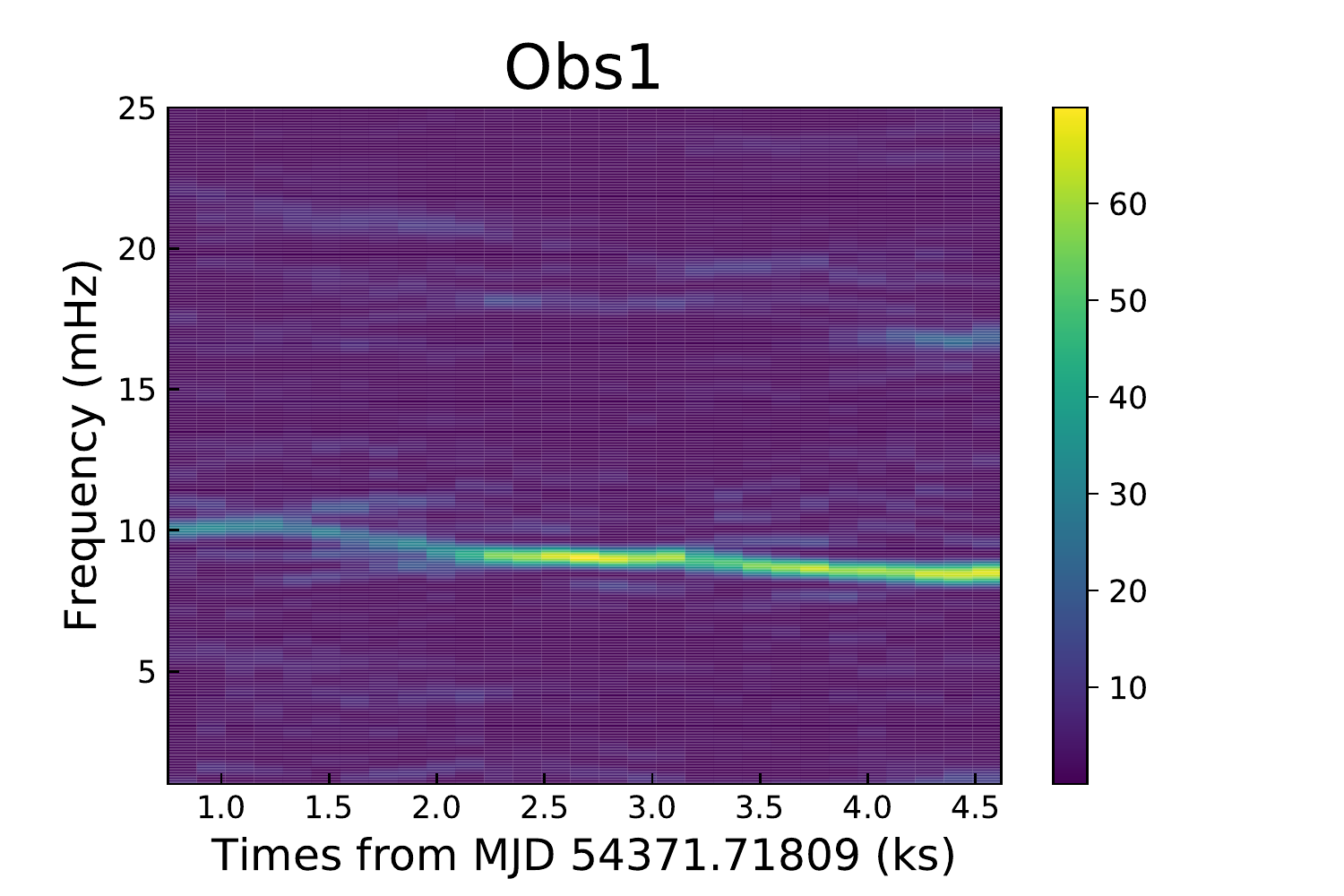}{0.5\textwidth}{}
          \fig{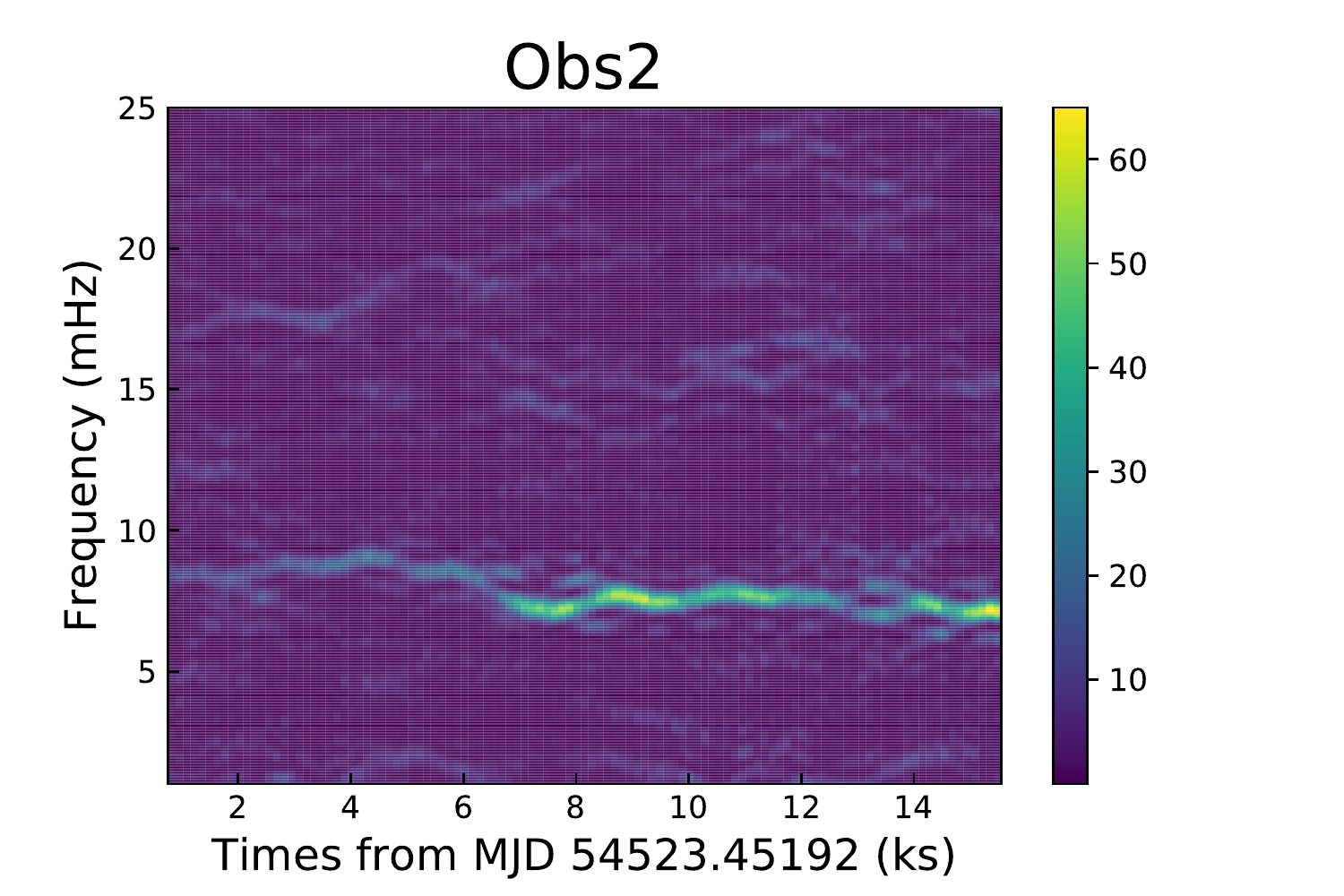}{0.5\textwidth}{}
          }
\gridline{\fig{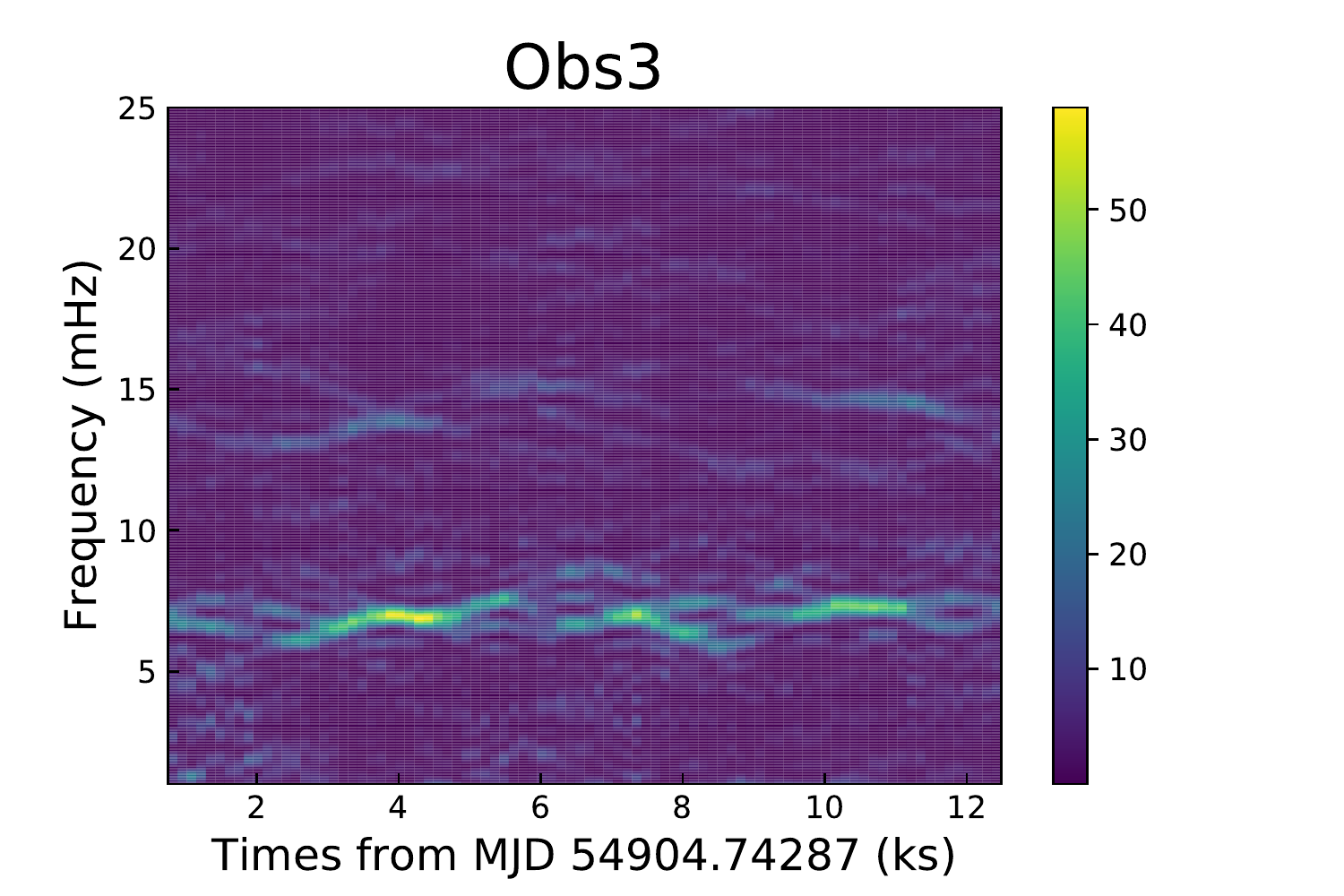}{0.5\textwidth}{}
          \fig{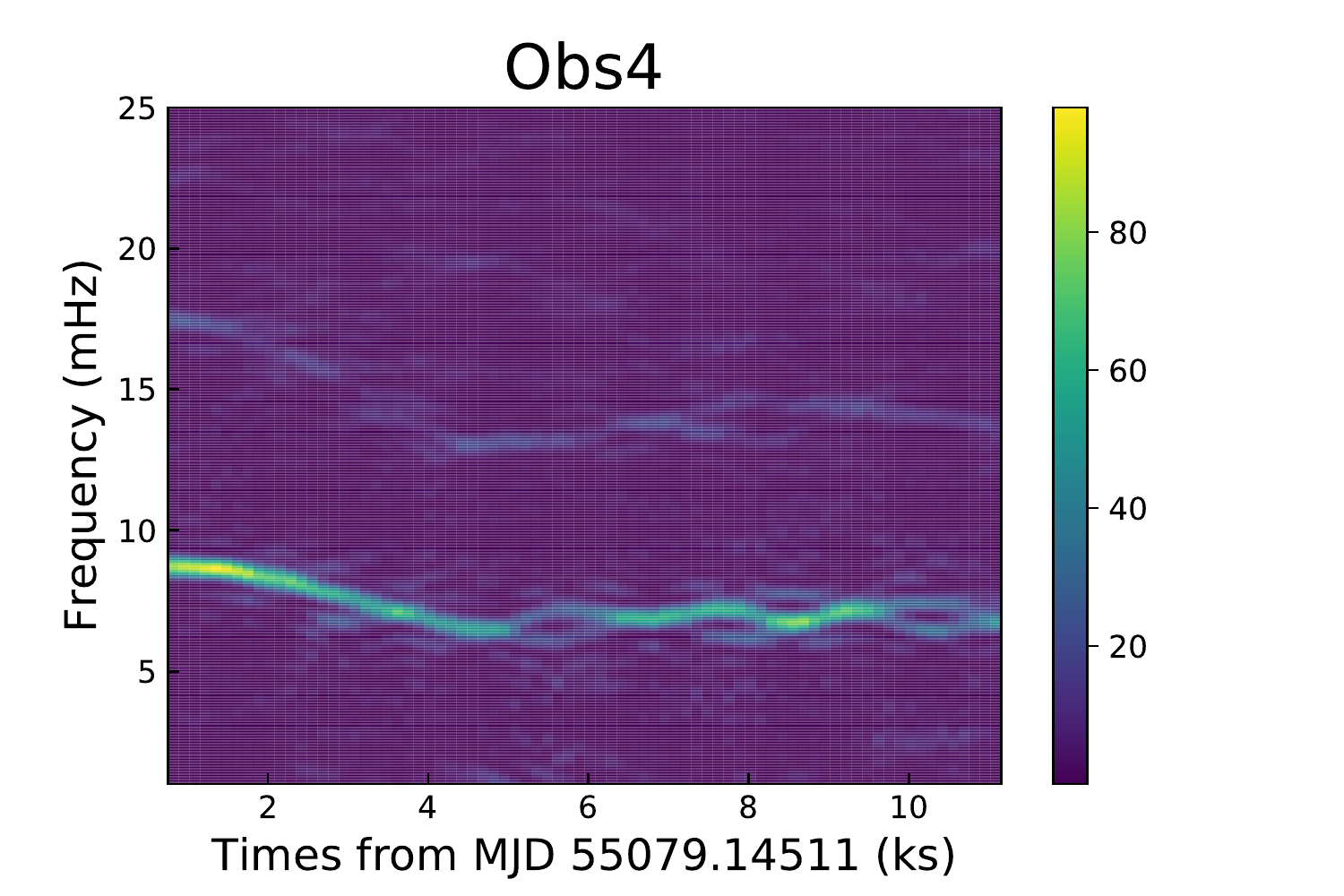}{0.5\textwidth}{}
          }
\caption{Dynamic power spectra of the 4 observations with significant detections of mHz QPO. The frequency of mHz QPO drifts from $\sim$10 to 8 mHz as it approaches the type-I burst, and the variation in oscillation power can be clearly seen.}
\label{fig1}
\end{figure*}

All photon arrival times were first corrected to the barycenter of the solar system by the SAS task {\tt barycen}. In order to optimize for the HHT analysis of the $\sim$8 mHz QPO, we binned the events collected in all 12 \emph{XMM-Newton} observations into 1.3333-second resolution light curves. The selection of this time resolution will be explained later in this section. The dynamic power spectral analysis was adopted for these 12 light curves to determine the time intervals with significant $\sim$8 mHz QPO detections. These power spectra were obtained by the Lomb-Scargle power periodogram \citep{1982ApJ...263..835S} with a window size of 1130 sec, $\sim$10 cycles of QPO variations, and moving step size of 100 sec. The $\sim$8 mHz QPO can be clearly seen only in 4 of 12 light curves, labelled as Obs1 to Obs4 (see Table \ref{tab:ID}), whose dynamic power spectra are shown as Figure \ref{fig1}. Except for Obs3, the frequency was decreased before a Type-I X-ray burst according to the dynamic power spectrum, consistent with the previous study \citep{2015MNRAS.454..541L}.

Although the variation in the frequency and power can be clearly seen, the resolution of the dynamic power spectrum is limited by the window size, that is, $\delta$f$\delta$T$\approx$1 ,where $\delta$f and $\delta$T are the resolutions of frequency and time (i.e.,\ window size), respectively. This limitation prevents us from further analysis of this variation. To investigate the $\sim$8 mHz QPO in more detail, the HHT was applied in the following of this work. However, a continuous light curve is preferred for the HHT analysis but there are gaps in Obs2 and Obs3. Therefore, we only analyzed the parts after the gaps because they are closer to the corresponding type-I bursts.

We performed HHT analysis for these 4 selected light curves. The HHT is a method for analyzing the nonlinear and non-stationary signal. This method consists of two major steps \citep{2008RvGeo..46.2006H}: (1) Using empirical mode decomposition (EMD) to adaptively decompose a time series into intrinsic mode functions (IMFs). (2) To obtain the instantaneous frequency and amplitude of the IMFs through the Hilbert transform. EMD is a method for decomposing any time series into several IMFs. This process was developed by \citet{1998RSPSA.454..903H} and then further improved by \citet{2009Huang..Wu}. IMFs are time series that are suitable for the Hilbert transform to resolve instantaneous frequency and amplitude of non-stationary periodic oscillations, such as QPOs.

In this work, we applied the fast-complementary ensemble empirical mode decomposition (CEEMD) method with post-processing  \citep{2009Huang..Wu,2010Yeh..Huang,2014PhyA..400..159W}, a modified version of EMD. EMD is a sifting process for separating oscillation modes from original data by subtracting the local means in the data \citep{1998RSPSA.454..903H}. These decomposed components are IMFs that satisfy the following conditions: (1) the number of extrema and the number of zero crossings must either be equal or differ at most by one, and (2) at any point, the mean value of the envelope defined by the local maxima and the envelope defined by the local minima is zero \citep{1998RSPSA.454..903H}. However, IMFs extracted by EMD may suffer from the mode mixing problem, in which a modulation with the same timescale is distributed across different IMFs \citep{2010Yeh..Huang}. CEEMD can overcome this mode mixing problem by (1) adding white noise to original data (2) decomposing the noisy data into IMFs by EMD (3) repeating (1) and (2) several times with different white noise each time (4) averaging these IMFs. Because CEEMD involves numerous summations of IMFs, it indicates that the CEEMD components of data may not be IMFs. In order to guarantee that the final result satisfied the IMF criteria, the post-processing EMD \citep{2009Huang..Wu} was applied to the decomposed components.

\begin{figure}[ht!]
\plotone{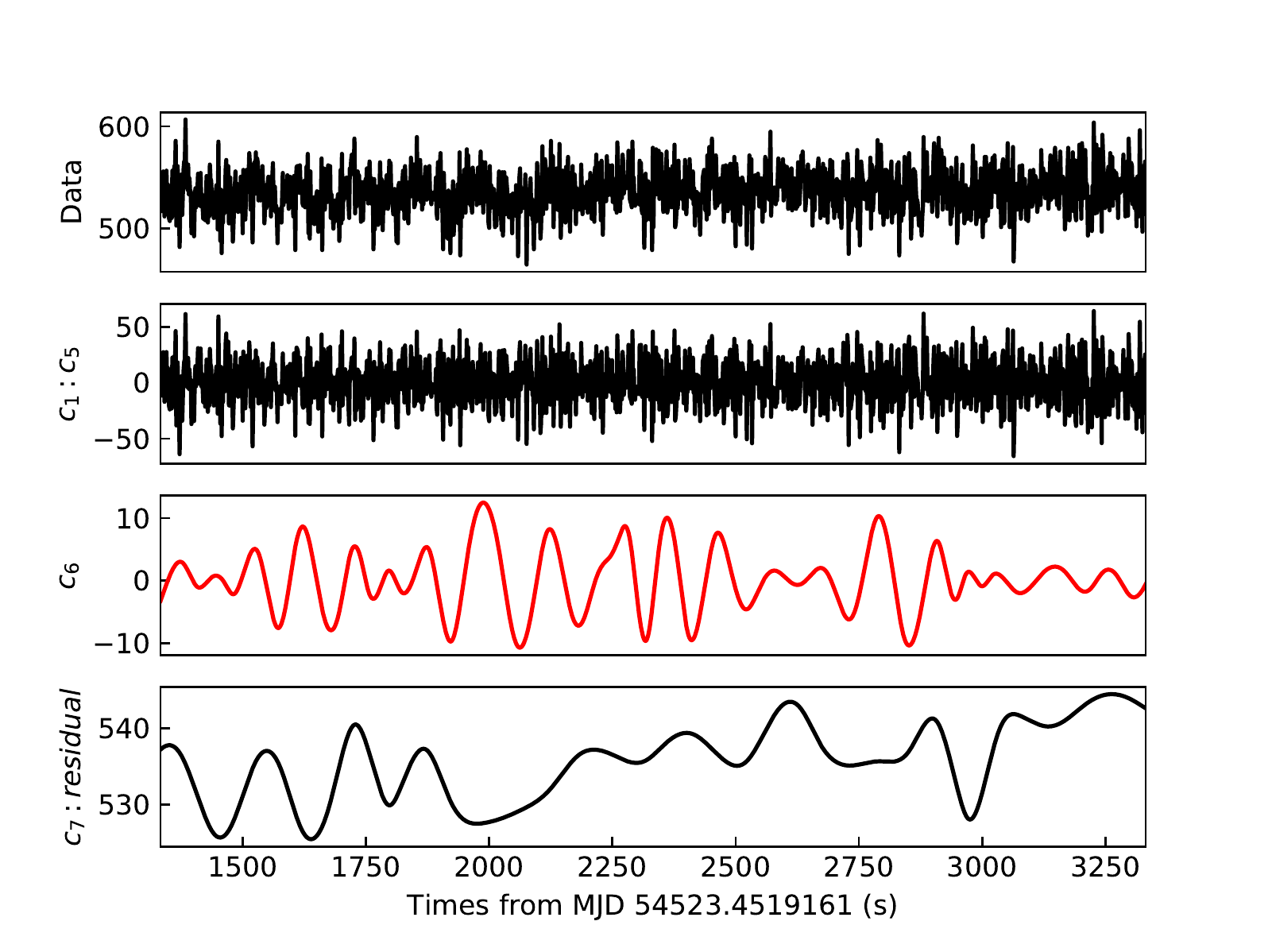}
\caption{Example of a 1500s light curve, which decomposes to IMFs. (a) The original light curve. (b) The high-frequency noise from the summation of IMF $c_{1}$ to $c_{5}$. (c) The IMF $c_{6}$, which corresponds to $\sim$8 mHz QPOs. (d) The low-frequency noise from the summation of IMF $c_{7}$ to the residual.}
\label{fig2}
\end{figure}

\begin{figure*}[ht!]
\gridline{\fig{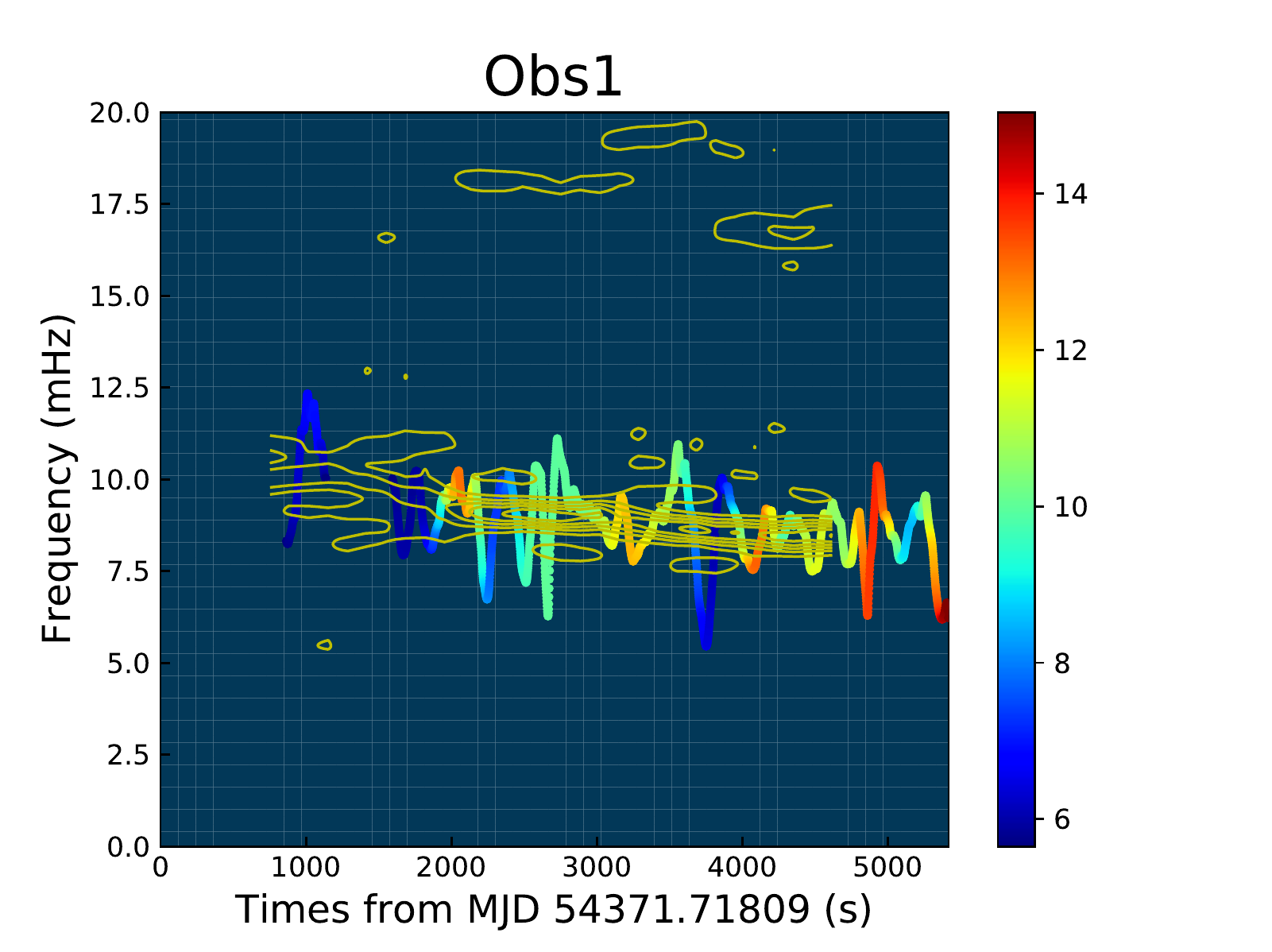}{0.5\textwidth}{}
          \fig{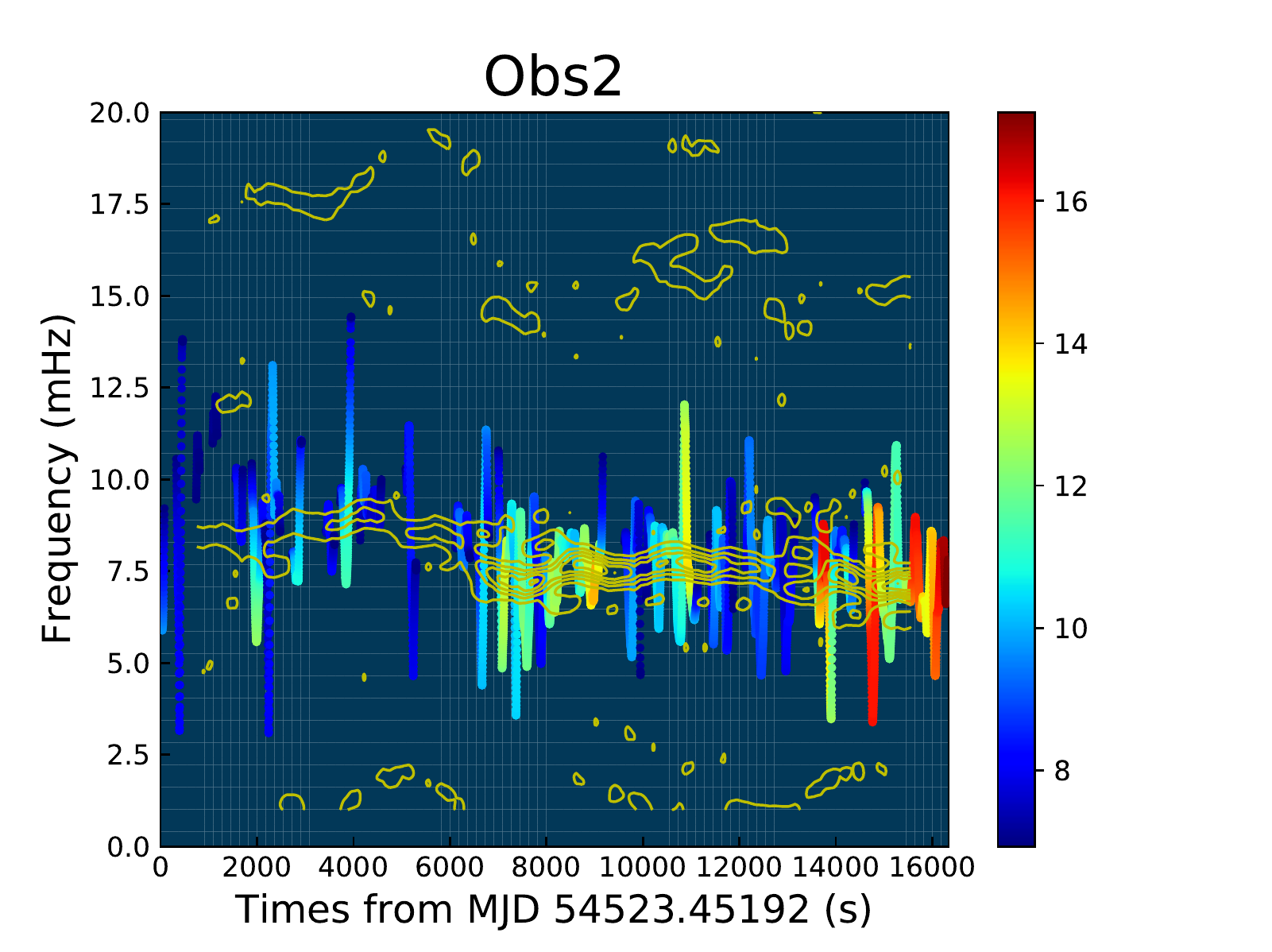}{0.5\textwidth}{}
          }
\gridline{\fig{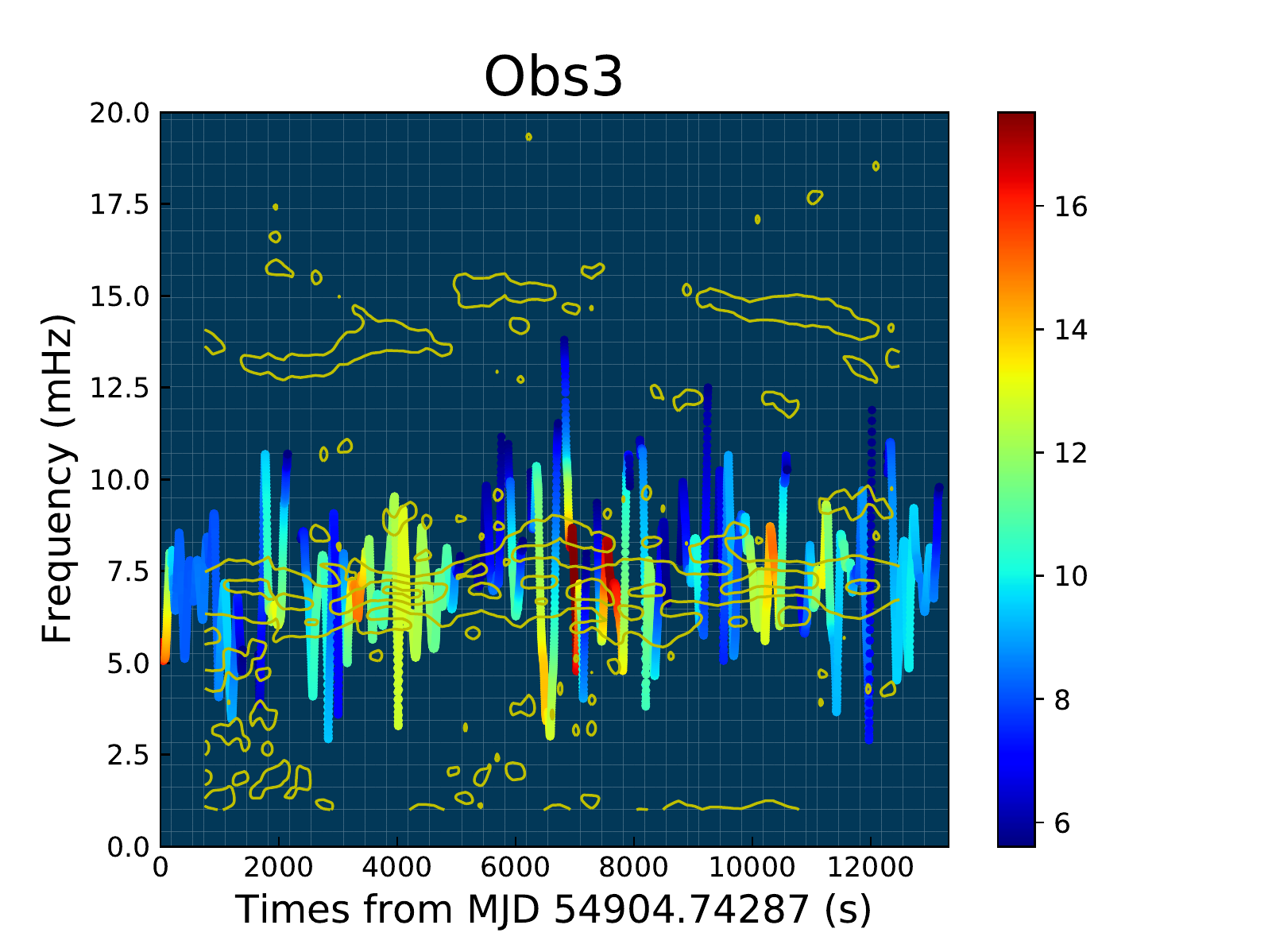}{0.5\textwidth}{}
          \fig{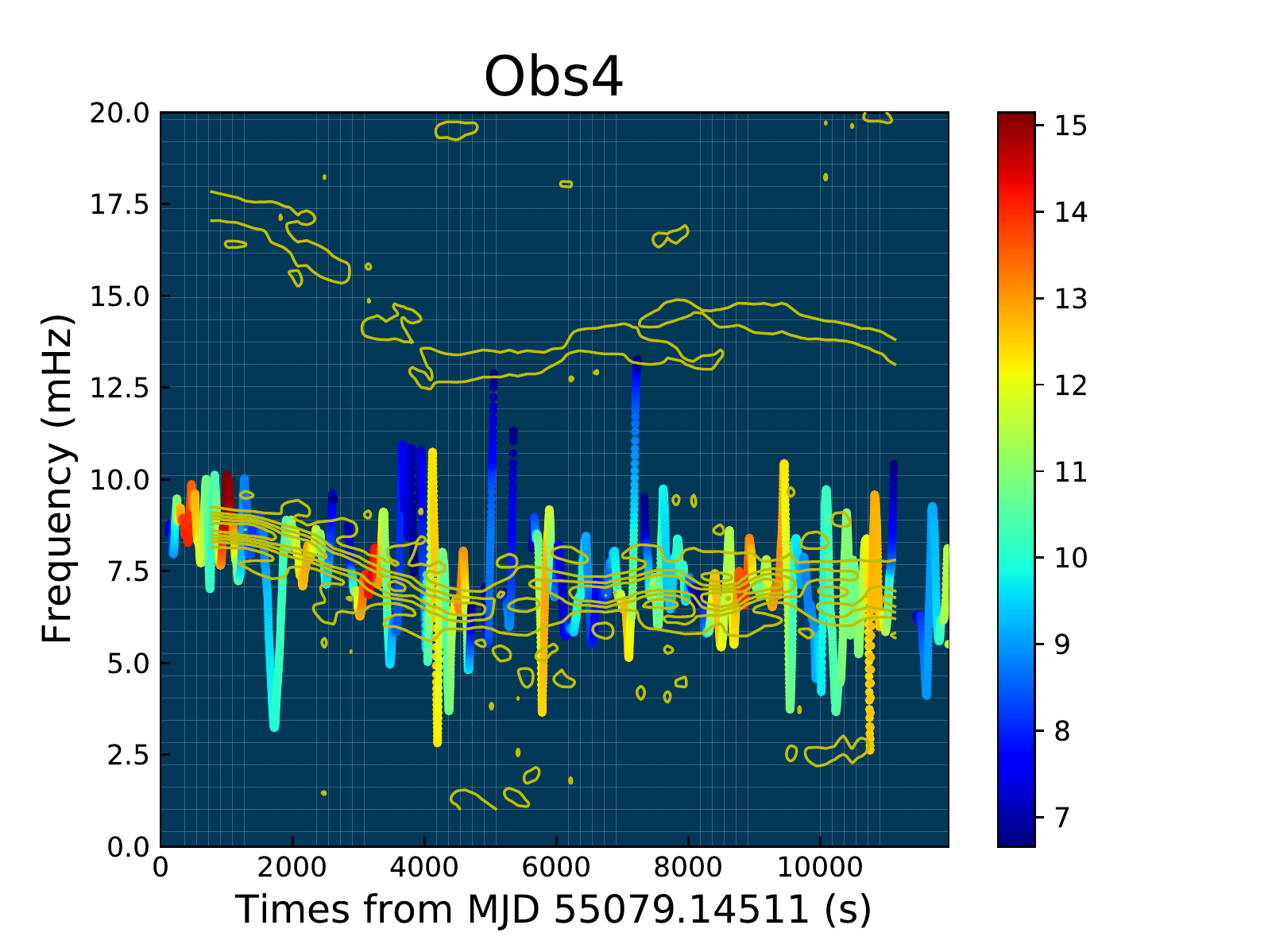}{0.5\textwidth}{}
          }
\caption{Hilbert spectra of mHz QPOs. The color on the z-axis represents the amplitude. Some instantaneous frequency drift above 15 mHz, but their amplitudes are less 3$\sigma$ significance ($\sim$6 cts/s). These data were neglected when we applied the confidence limits. The contour is the dynamic Lomb-Scargle periodogram of the detrended light curve.}
\label{fig3}
\end{figure*}

The light curve, $x(t)$, may be expressed as the sum of the IMFs, $c_{j}(t)$, and the residual $r_{n}(t)$, $x(t)=\sum_{j=1}^{n}c_{j}(t)+r_{n}(t)$, where n is the number of IMFs. Because the variation time scale of the first IMF component is approximately 3 data points, i.e.,\ 3 $\times$ 1.3333s = 4s, and the variation in the time scale of the second component is twice of the first component \citep{1998RSPSA.454..903H} and so on, the time scale in which we are interested ($\sim$125s) is expected to concentrate at the $6^{th}$ IMF component ($4 \times 2^{(6-1)} = 128s$). After decomposing the light curve, we determined that the $\sim$8 mHz oscillation can be observed in the sixth IMF, that is, $c_{6}(t)$ as expected (see Figure \ref{fig2}), by confirming the orthogonality of the IMF components \citep{1998RSPSA.454..903H,2014ApJ...788...31H}. We then applied the normalized Hilbert transform \citep{2009Huang..Wu} to obtain the instantaneous phase, frequency and amplitude of IMF $c_{6}(t)$. The instantaneous amplitude $a_{6}(t)$ is defined as the cubic Hermit spline envelope of the local maxima of the absolute values of the IMF $c_{6}(t)$ \citep{2009Huang..Wu}. The Hilbert transformation of the normalized IMF $X_{6}(t) = c_{6}(t) / a_{6}(t)$ can be represented as
\begin{eqnarray}
Y_{6}(t)=\frac{1}{\pi}P\int_{-\infty}^{\infty}\frac{X_{6}(t')}{t-t'}dt' \label{eq1}
\end{eqnarray}
where P is the Cauchy principal value. Thus, we can define an analytical signal $Z_{6}(t)$ and the instantaneous phase function $\theta_{6}(t)$ as
\begin{eqnarray}
Z_{6}(t)=X_{6}(t)+iY_{6}(t)= e^{i\theta_{6}(t)} \label{eq2}
\end{eqnarray}
Therefore, the instantaneous frequency $\nu_{6}(t)$ can be defined as
\begin{eqnarray}
\nu_{6}(t)=\frac{1}{2\pi}\frac{d\theta_{6}(t)}{dt} \label{eq3}
\end{eqnarray}
Figure \ref{fig3} shows Hilbert spectra of  $\sim$8 mHz oscillation with more detailed variations in both frequency and amplitude.

To significantly improve detection, the confidence limits can be determined by repeating the CEEMD $10^{3}$ times with different sets of white noise to generate a thousand IMF $c_{6}(t)$, and then the means and standard deviations of the instantaneous amplitude and frequency from these IMF sets can be calculated. Finally, the significant signal is defined by the amplitude above the average of the $3\sigma$ lower limit ($\sim6$ cts/s) (see \citealt{2015ApJ...815...74S} for details) that the corresponding data will be selected for further analysis.

\subsection{Oscillation profile and phase resolved spectra} \label{subsec:phase_resolved}

\begin{figure*}[ht!]
\centering
\includegraphics[width=0.3\linewidth]{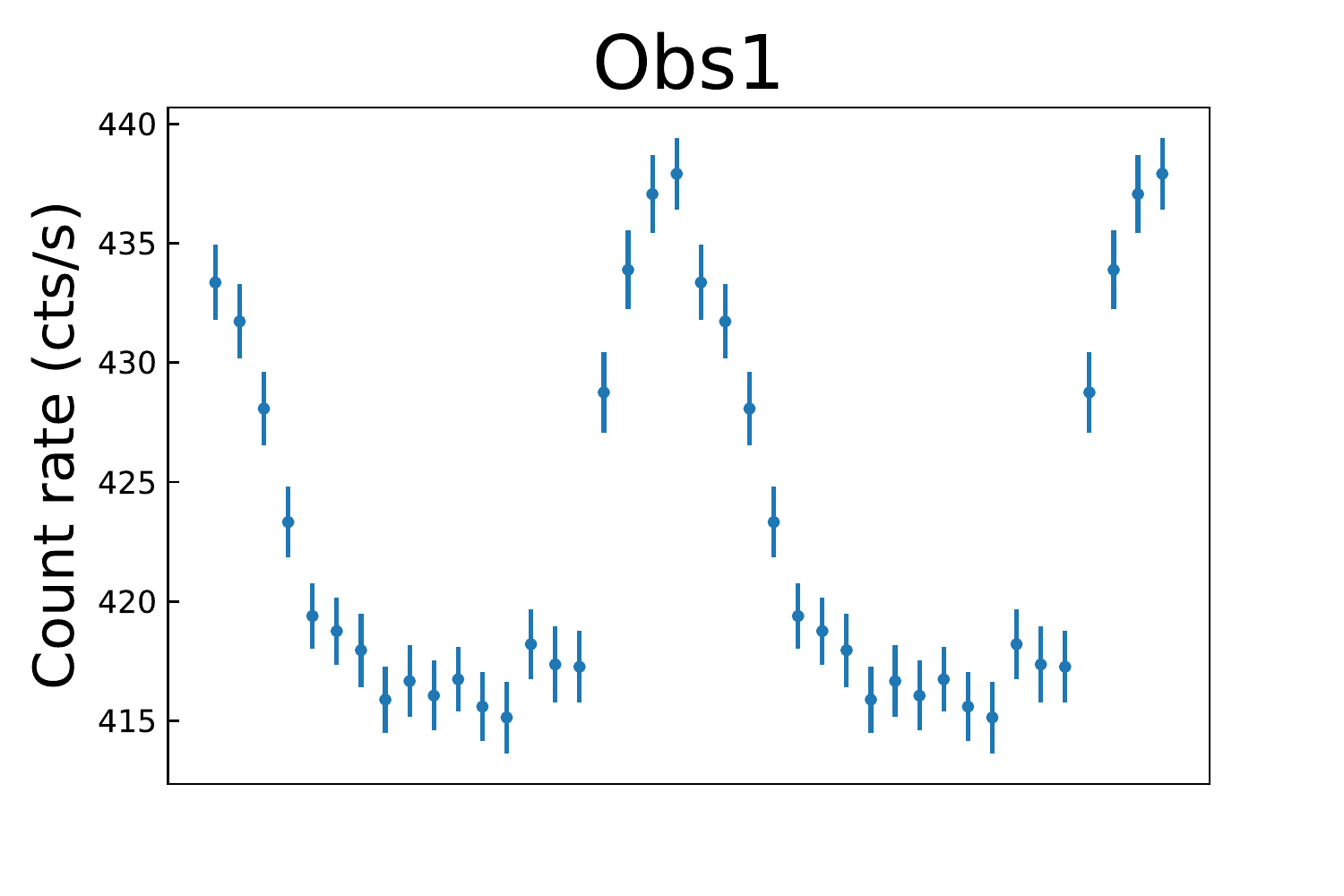}
\includegraphics[width=0.3\linewidth]{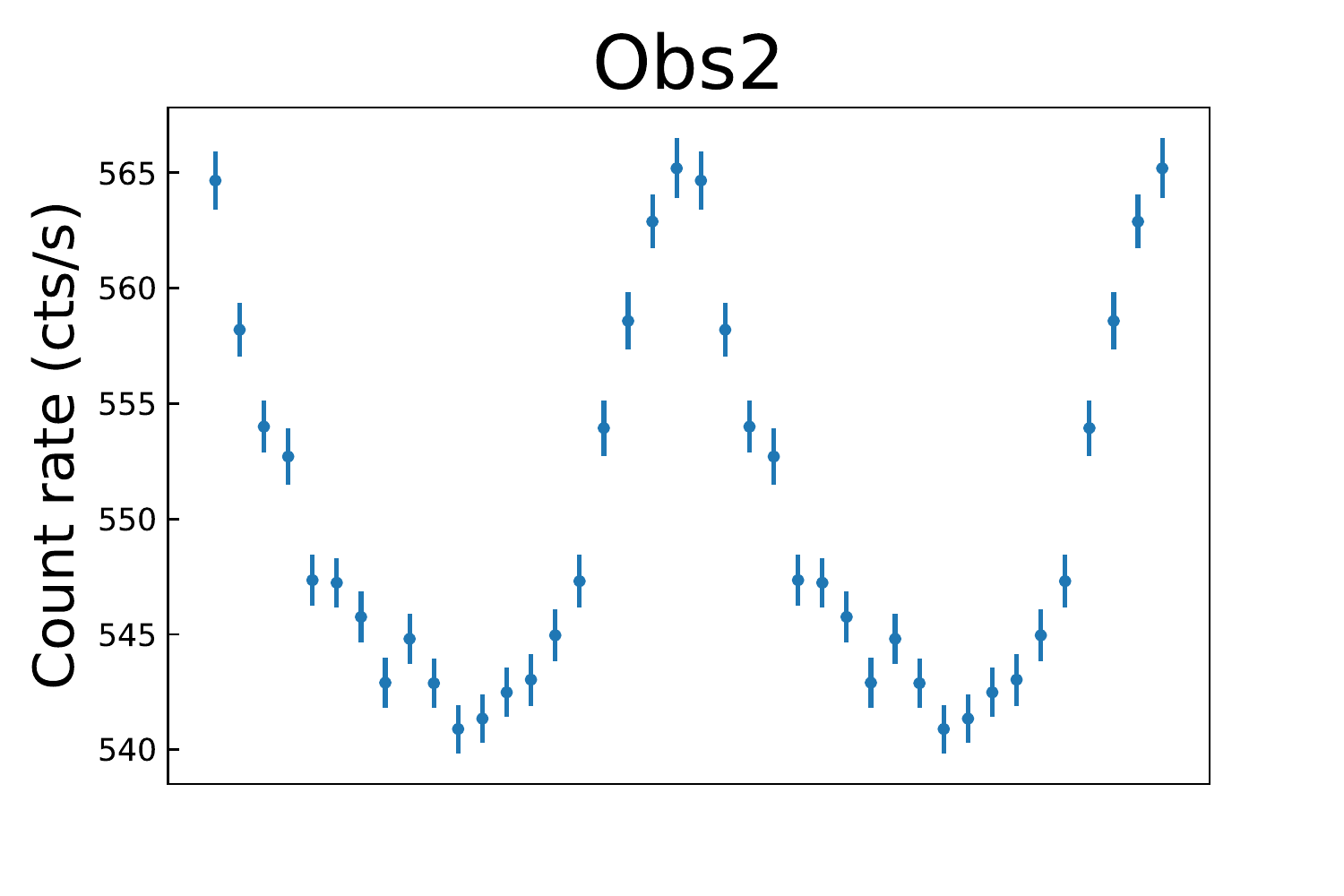}
\\
\includegraphics[width=0.3\linewidth]{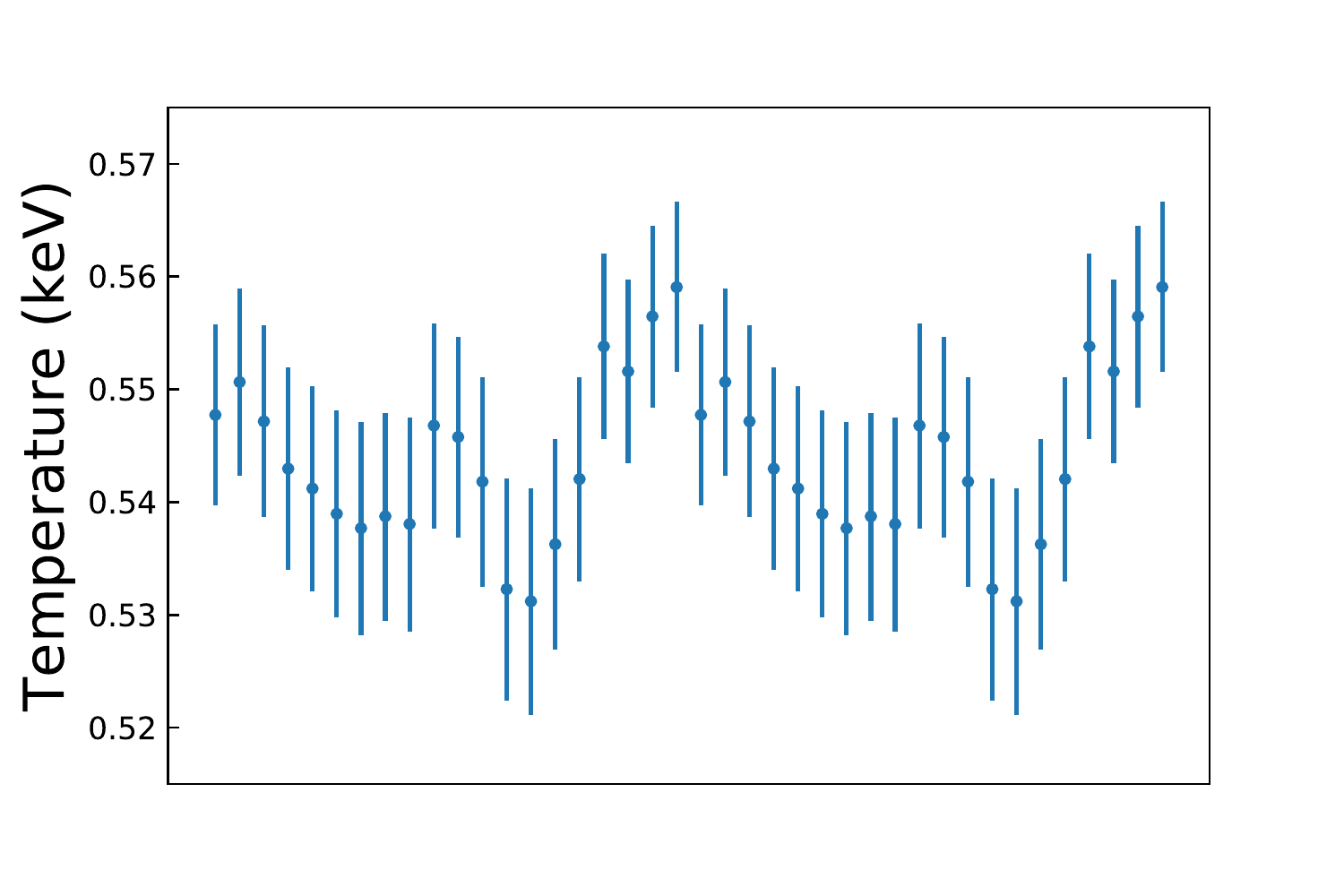}
\includegraphics[width=0.3\linewidth]{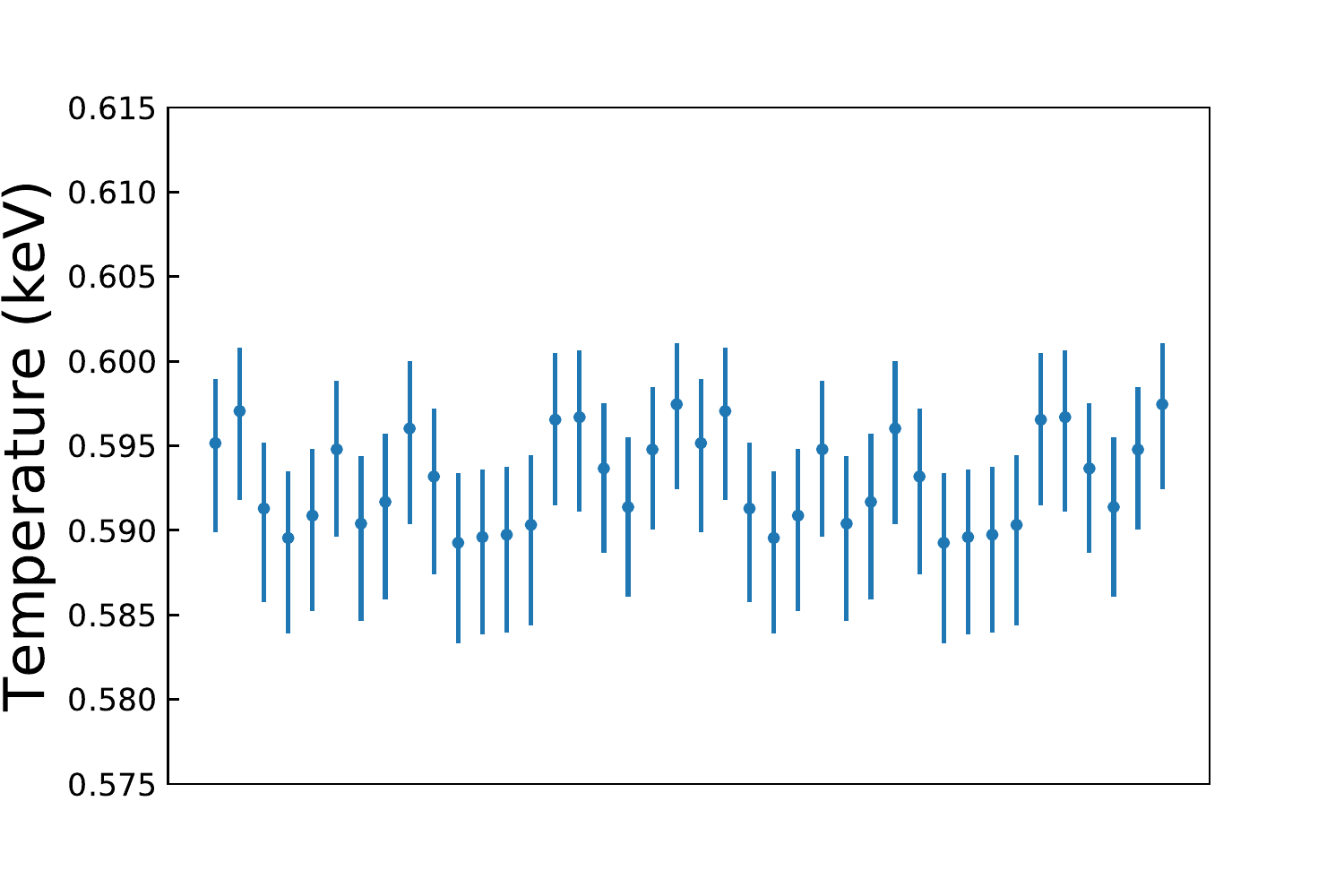}
\\
\includegraphics[width=0.3\linewidth]{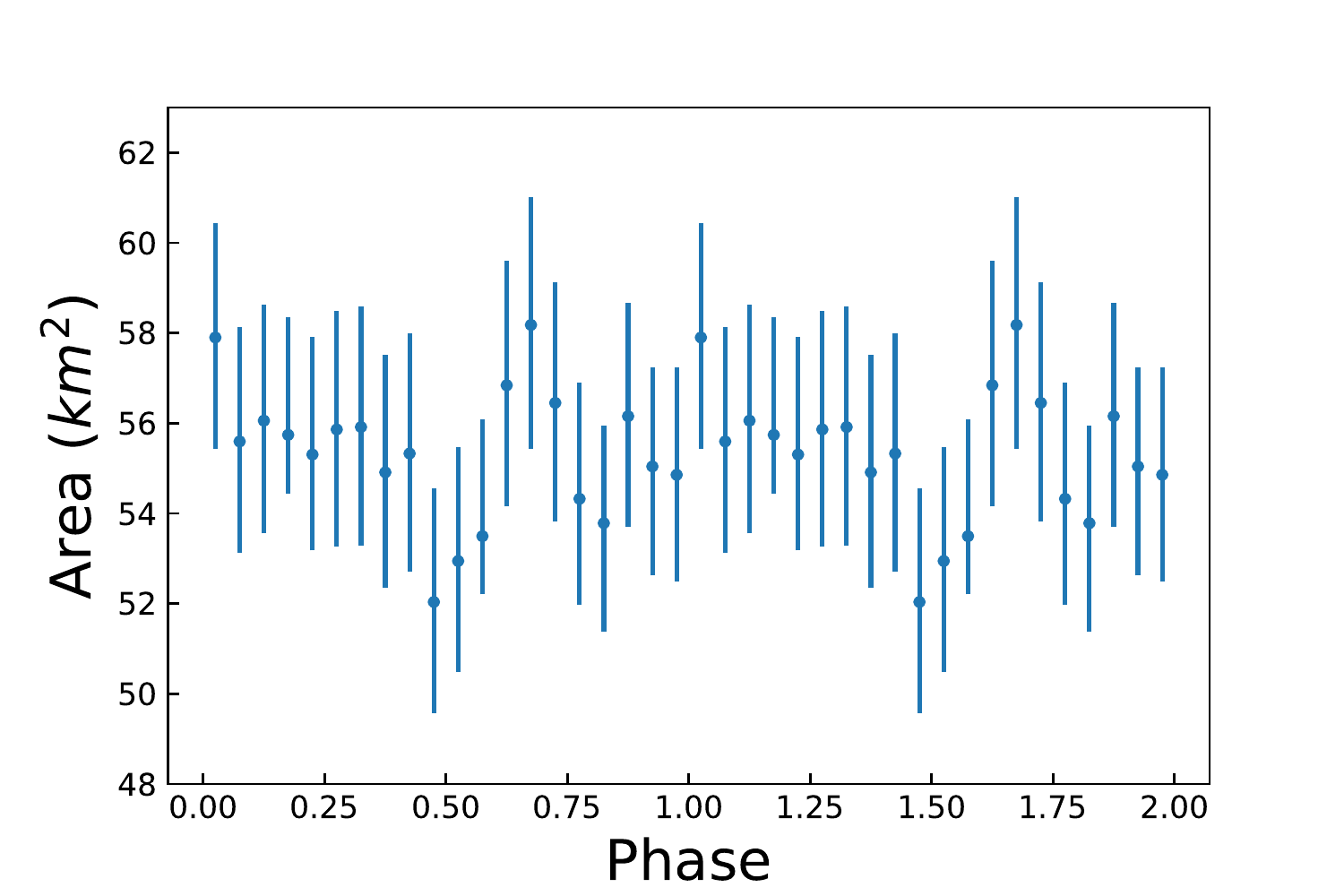}
\includegraphics[width=0.3\linewidth]{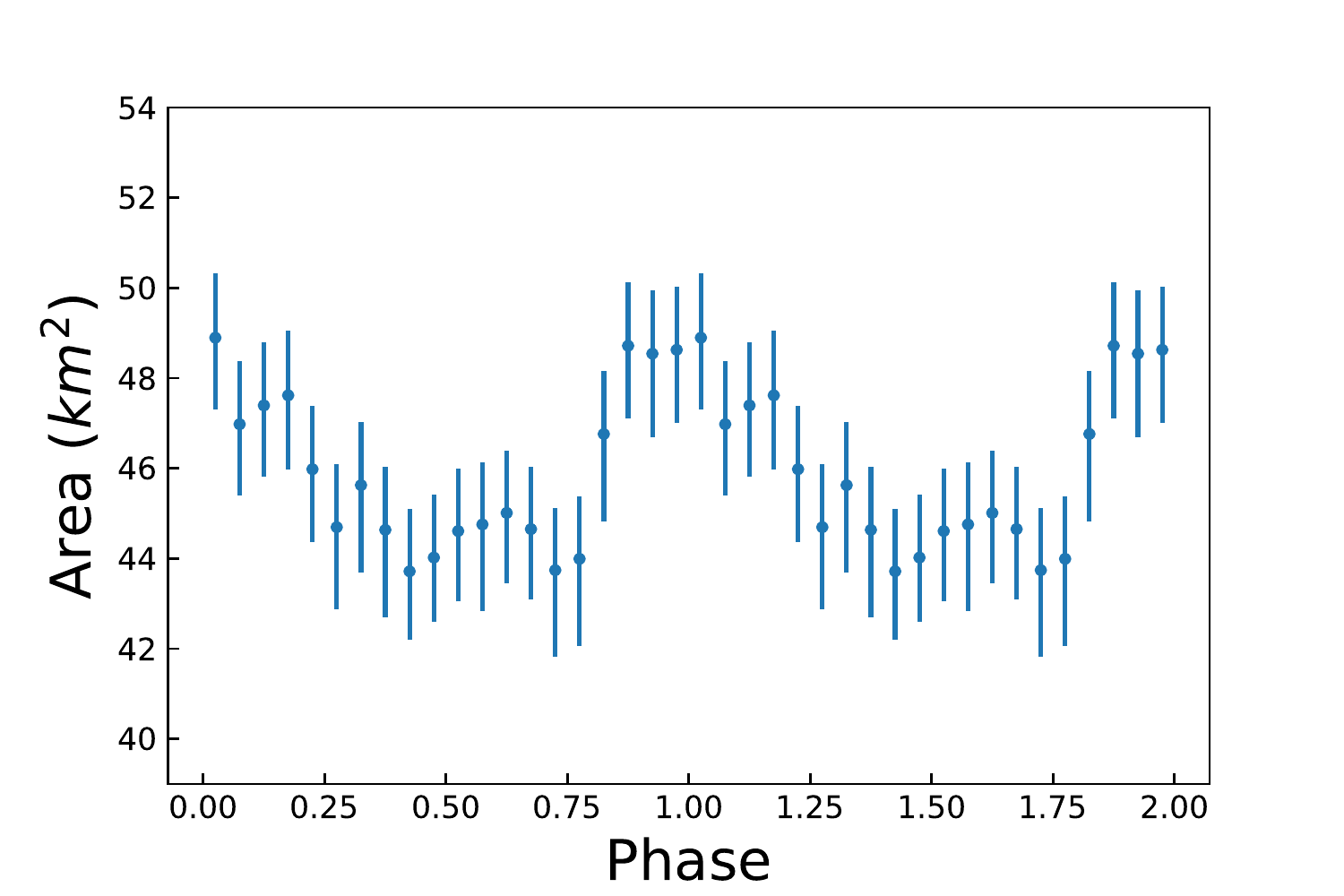}
\\
\includegraphics[width=0.3\linewidth]{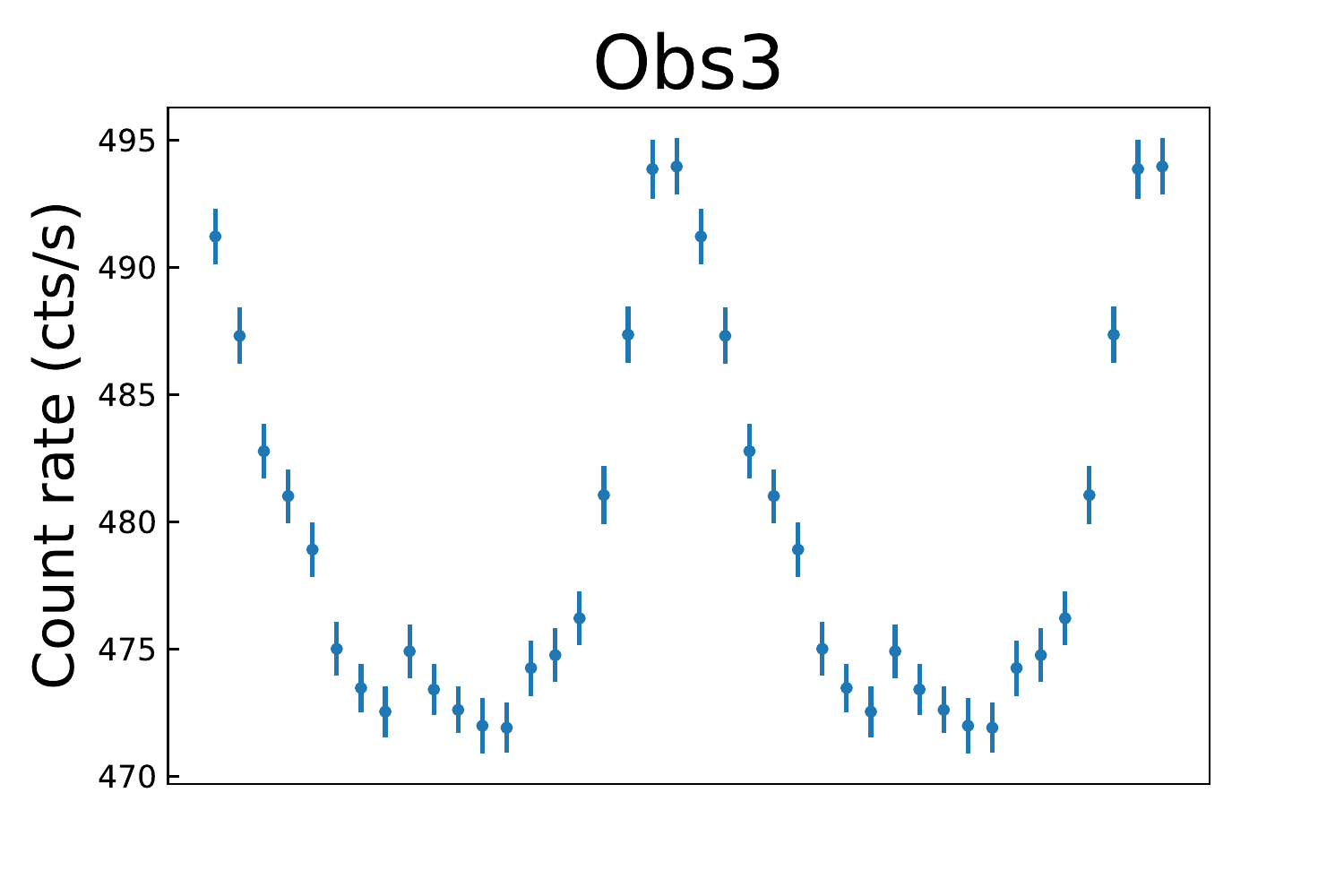}
\includegraphics[width=0.3\linewidth]{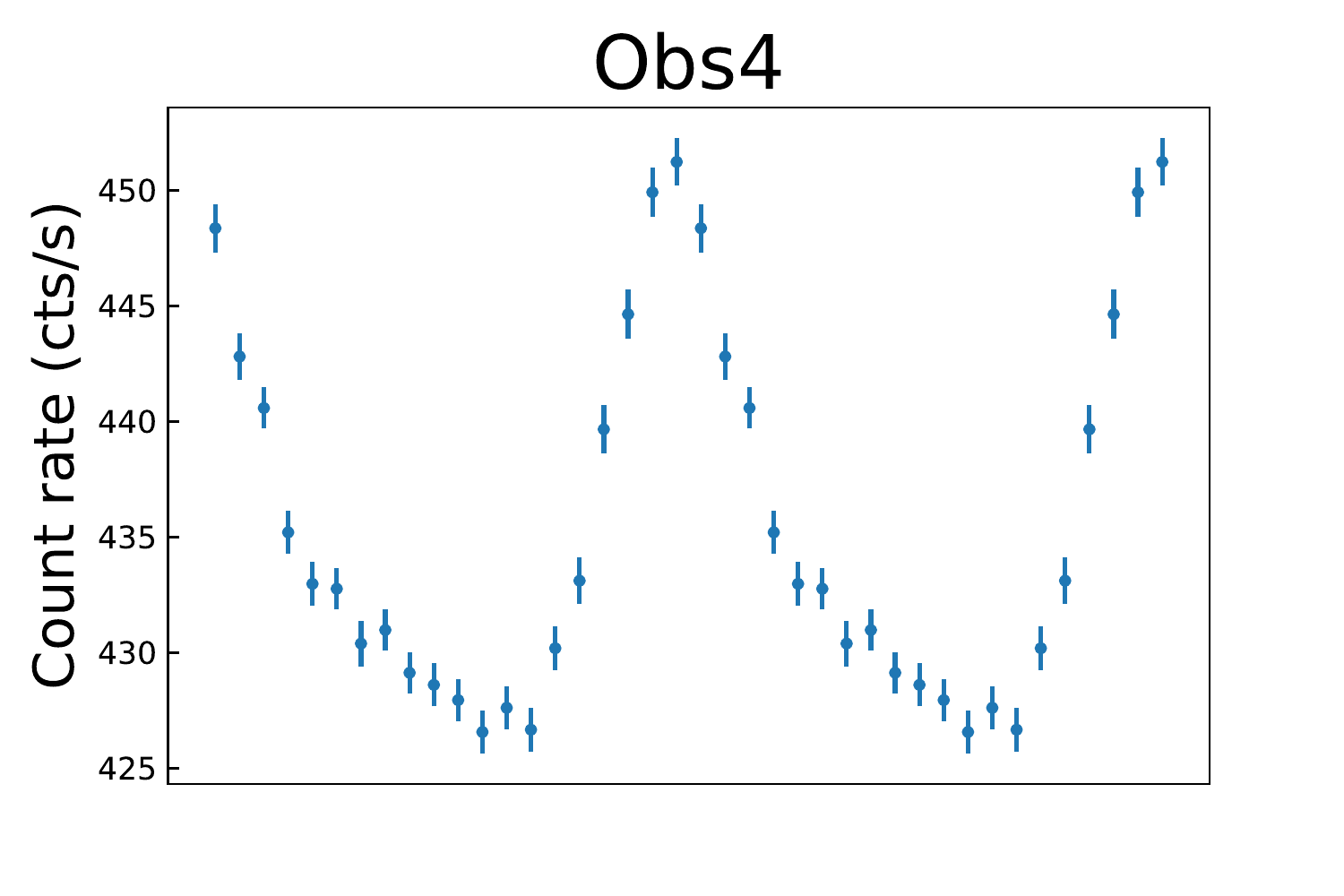}
\\
\includegraphics[width=0.3\linewidth]{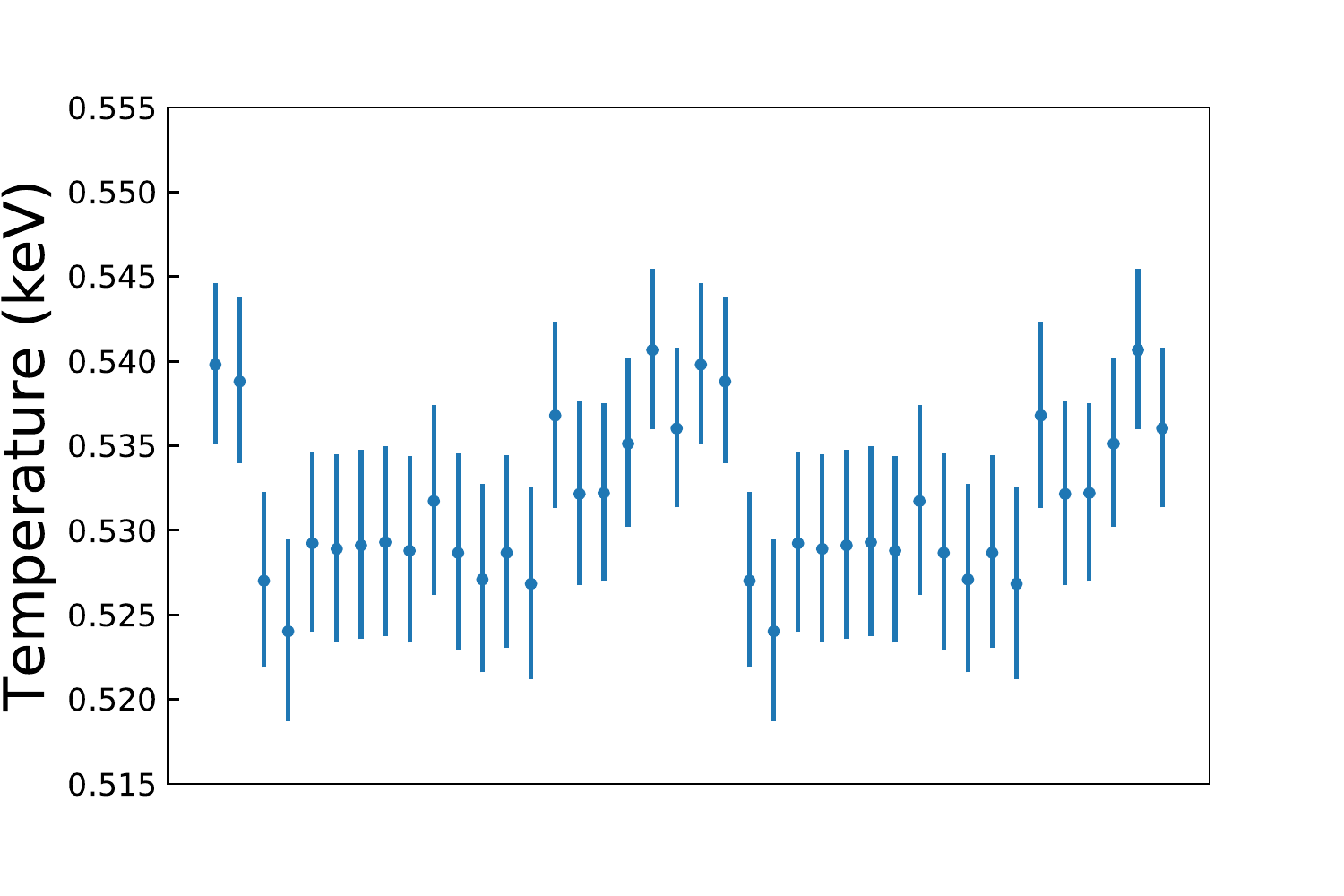}
\includegraphics[width=0.3\linewidth]{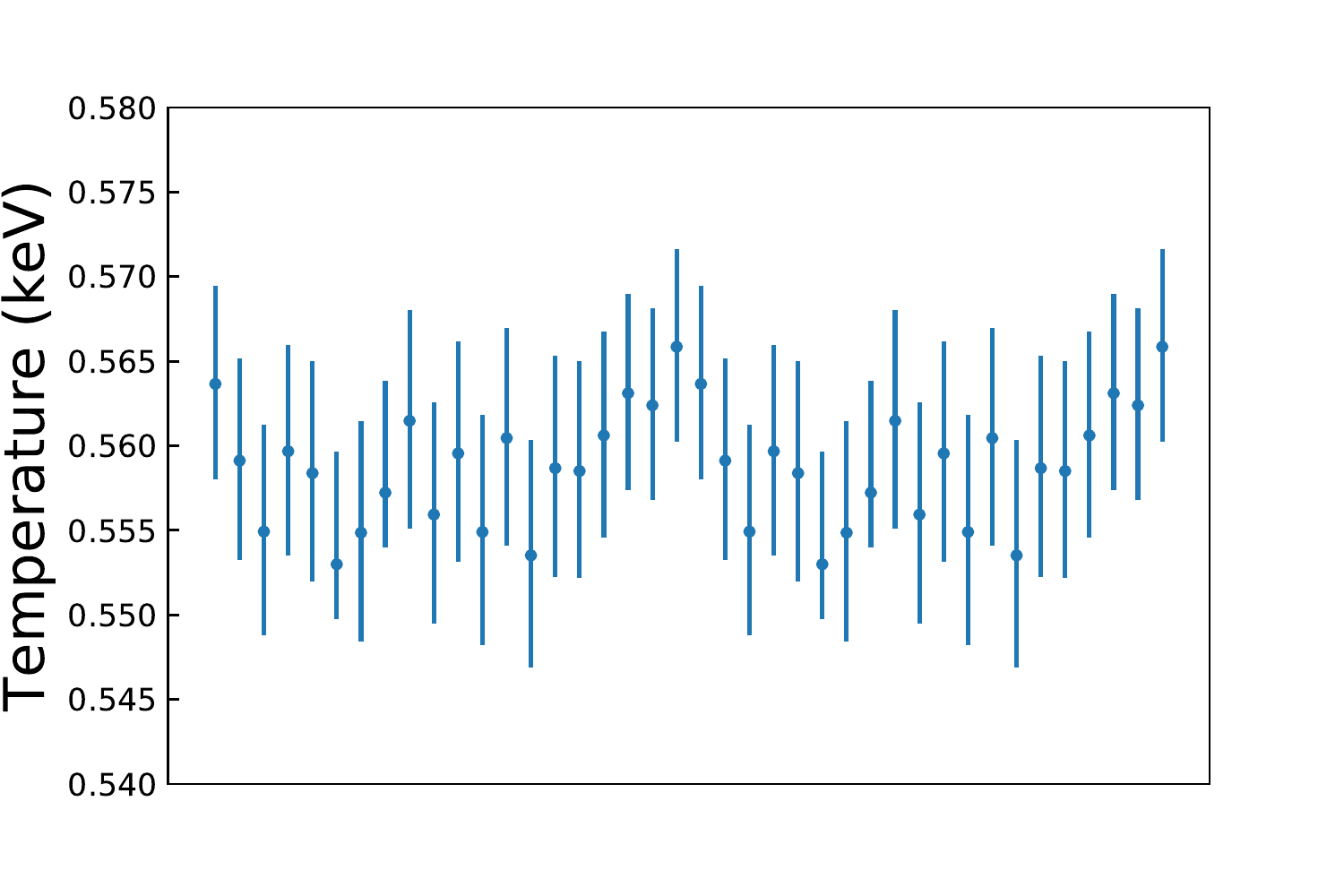}
\\
\includegraphics[width=0.3\linewidth]{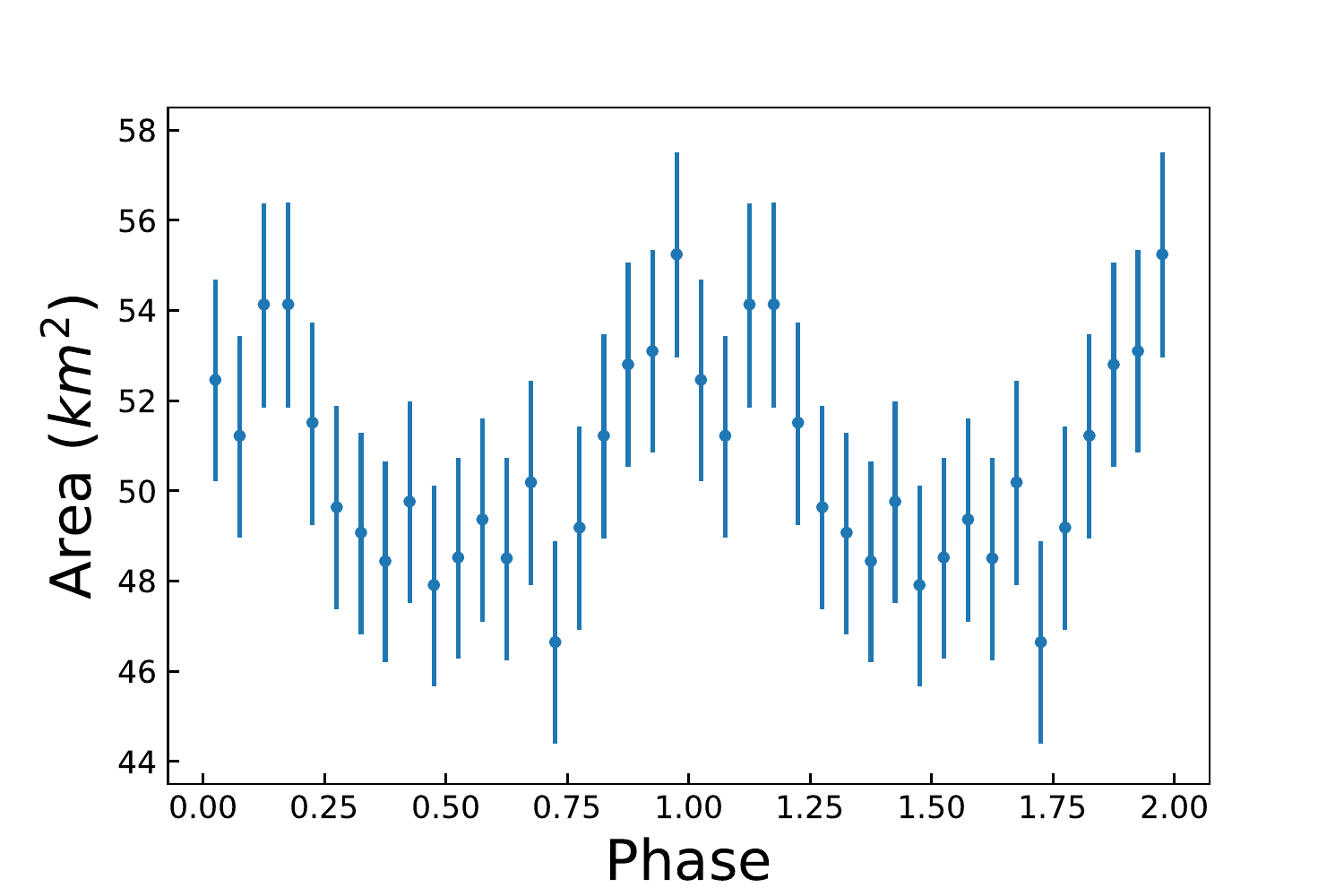}
\includegraphics[width=0.3\linewidth]{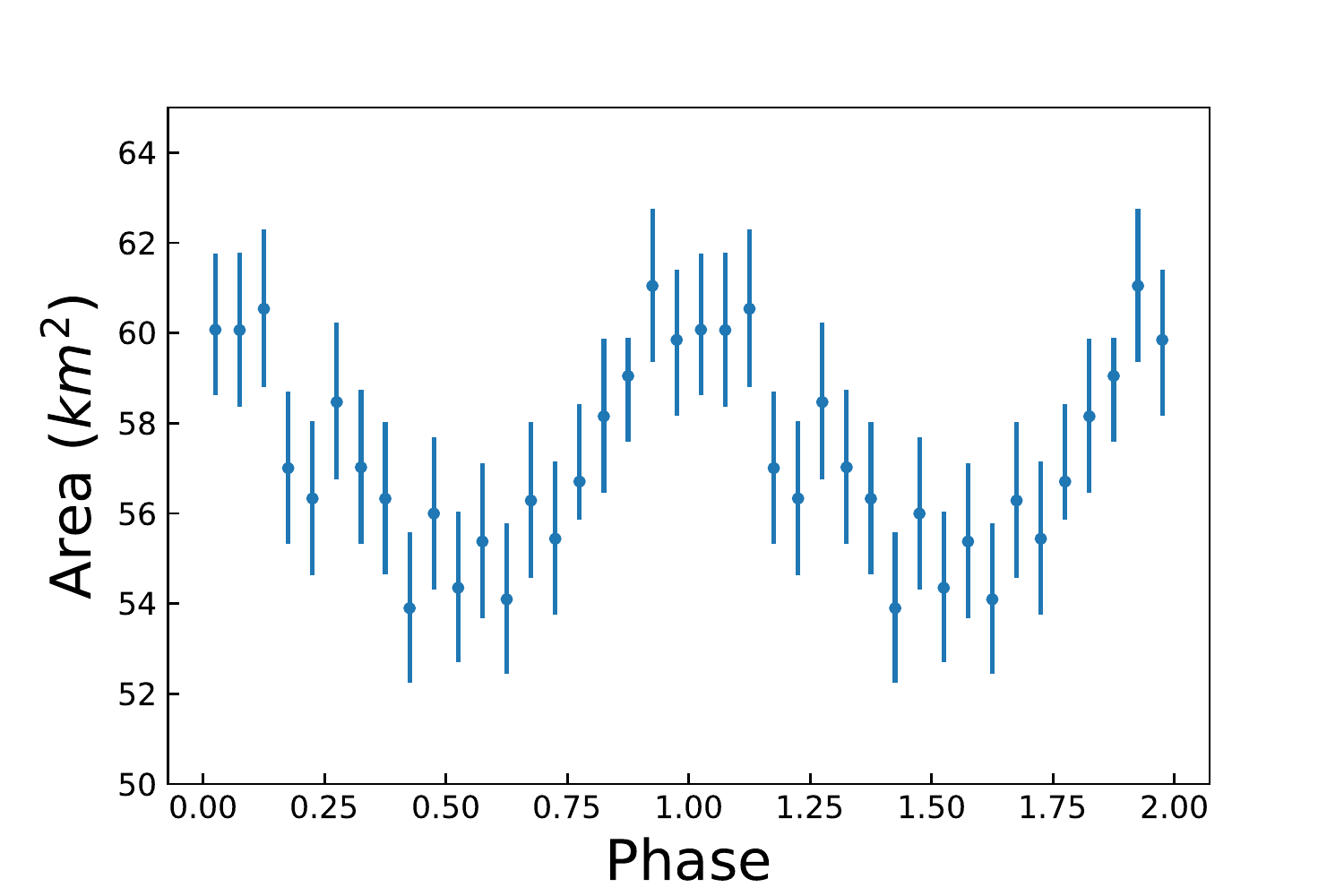}
%\gridline{\fig{L1_count.pdf}{0.4\textwidth}{}
%          \fig{L2_count.pdf}{0.4\textwidth}{}
%          }
%\gridline{\fig{L1_tem.pdf}{0.4\textwidth}{}
%          \fig{L2_tem.pdf}{0.4\textwidth}{}
%          }
%\gridline{\fig{L1_area.pdf}{0.4\textwidth}{}
%          \fig{L2_area.pdf}{0.4\textwidth}{}
%          }
%\gridline{\fig{L3_count.pdf}{0.4\textwidth}{}
%          \fig{L4_count.pdf}{0.4\textwidth}{}
%          }
%\gridline{\fig{L3_tem.pdf}{0.4\textwidth}{}
%          \fig{L4_tem.pdf}{0.4\textwidth}{}
%          }
%\gridline{\fig{L3_area.pdf}{0.4\textwidth}{}
%          \fig{L4_area.pdf}{0.4\textwidth}{}
%          }
\caption{Spectral parameter modulations of neutron-star surface temperature and apparent area. The neutron-star surface temperature is positively correlated with the variation in the neutron-star luminosity for Obs1. The apparent area is also positively correlated with the variation in the neutron-star luminosity for Obs2, Obs3 and Obs4.}
\label{fig4}
\end{figure*}

After the instantaneous QPO phases $\phi(t)=frac[\theta_{6}(t)]$ are evaluated by HHT (Eq.\ref{eq2}), the oscillation profile can be constructed by folding them even though the oscillation period is unstable. The oscillation profiles were constructed by binning the phases into 20 bins per cycle, as shown in Figure \ref{fig4}. The non-sinusoidal nature can be clearly seen in the oscillation profile, that may be approximately described as a fast rise and exponentially decay variation with the rising phase of $\sim$0.3 cycle and the decay phase of $\sim$0.5 cycle.

With the well-defined phase (Eq.\ref{eq2}), phase-resolved spectral analysis can be processed. We employed {\tt evselect}, a standard SAS tool, to extract energy spectra. The spectra were extracted by the selection criteria that PATTERN $<=$ 4 and FLAG $=$ 0. Because the source is very bright, the whole CCD was contaminated by source photons \citep{2014MNRAS.445.3659L,2015MNRAS.454..541L}. The background spectra were evaluated by the black hole candidate GX339-4 in quiescence state, that is in the similar sky region and column density along the line of sight \citep{2011MNRAS.411..137H, 2013MNRAS.432.1144S}.

\begin{figure*}[ht!]
\centering
\includegraphics[width=.3\textwidth, angle=270]{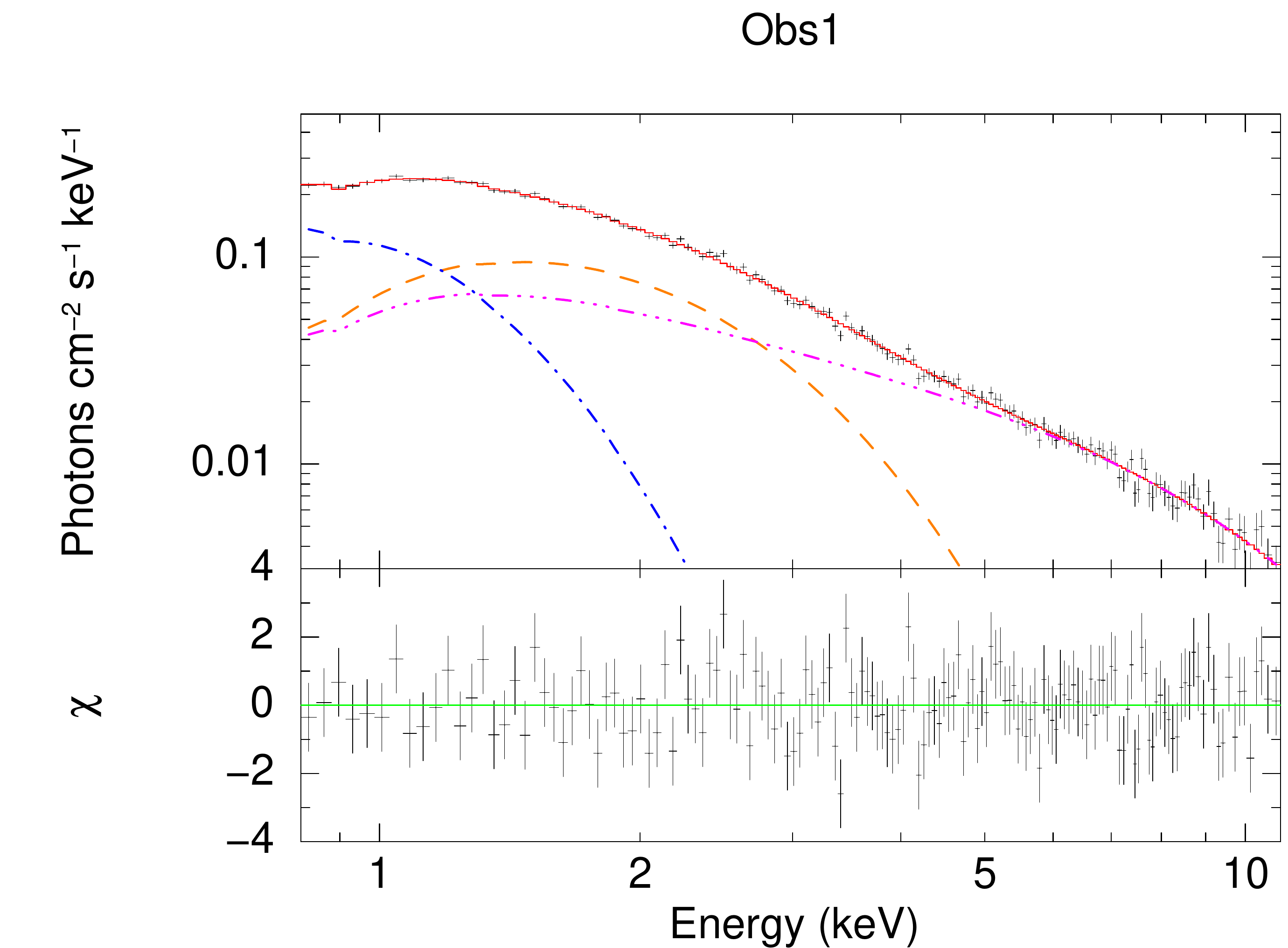}
\includegraphics[width=.3\textwidth, angle=270]{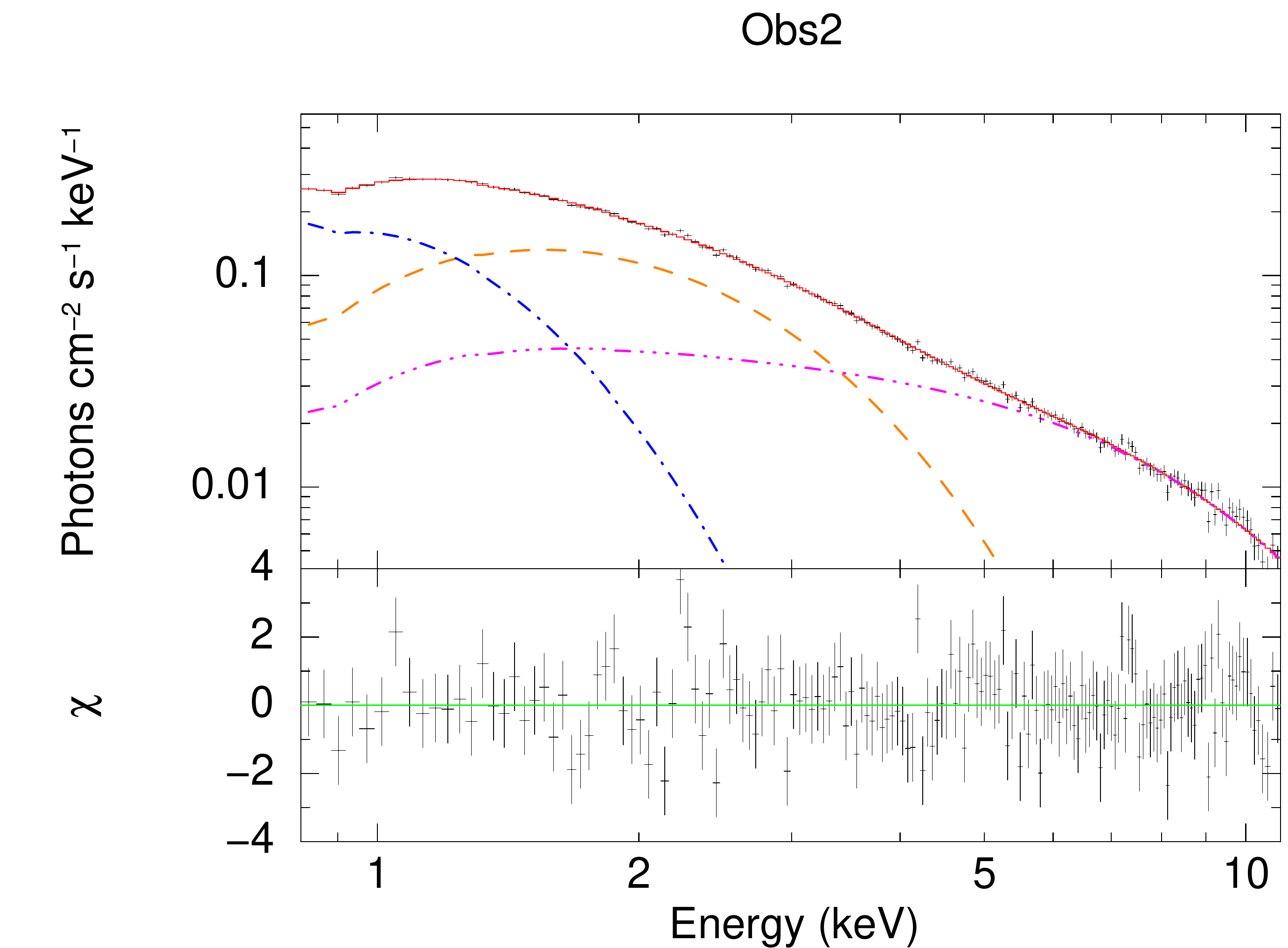}
\\
\vspace{0.4cm}
\includegraphics[width=.3\textwidth, angle=270]{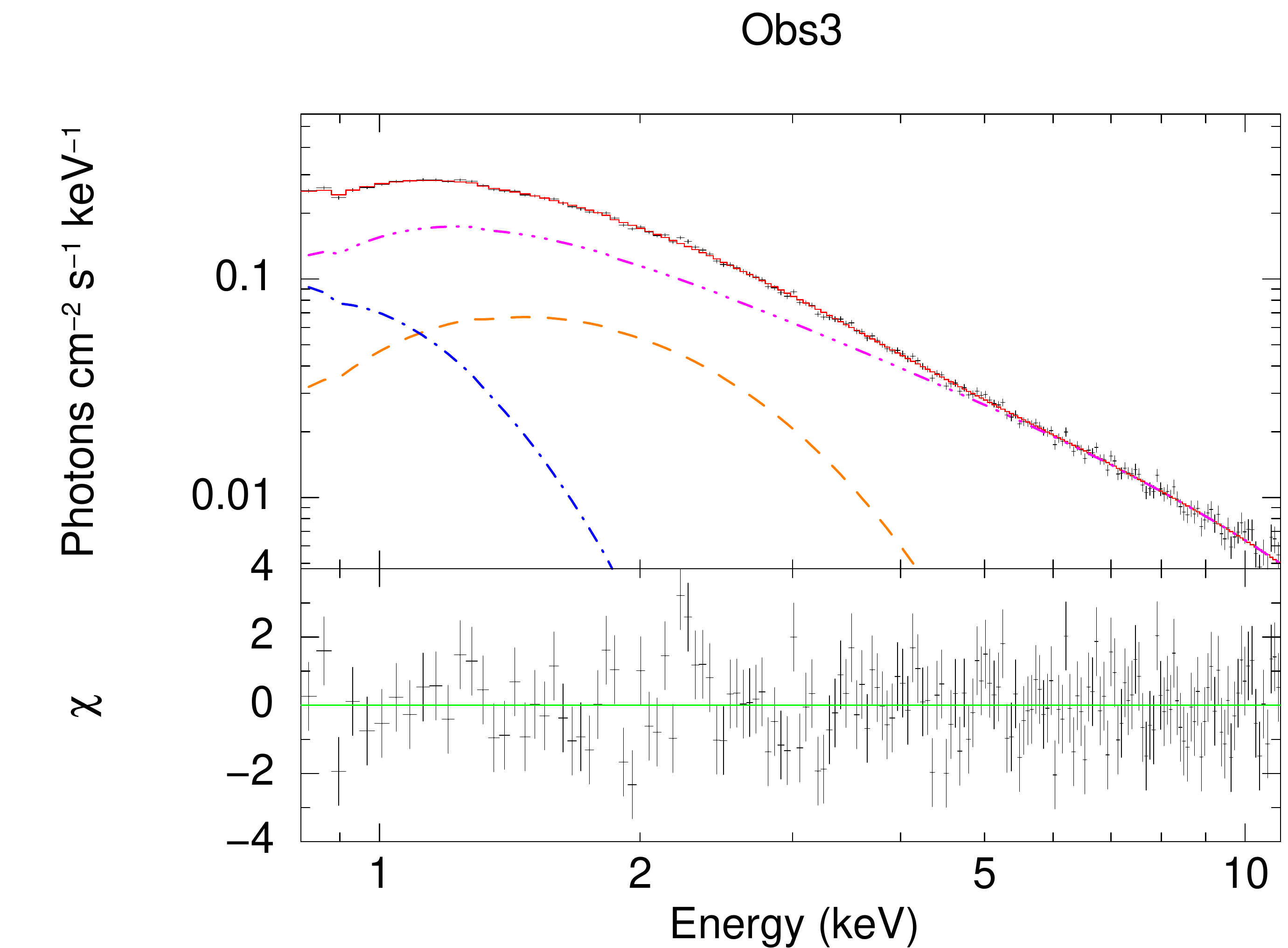}
\includegraphics[width=.3\textwidth, angle=270]{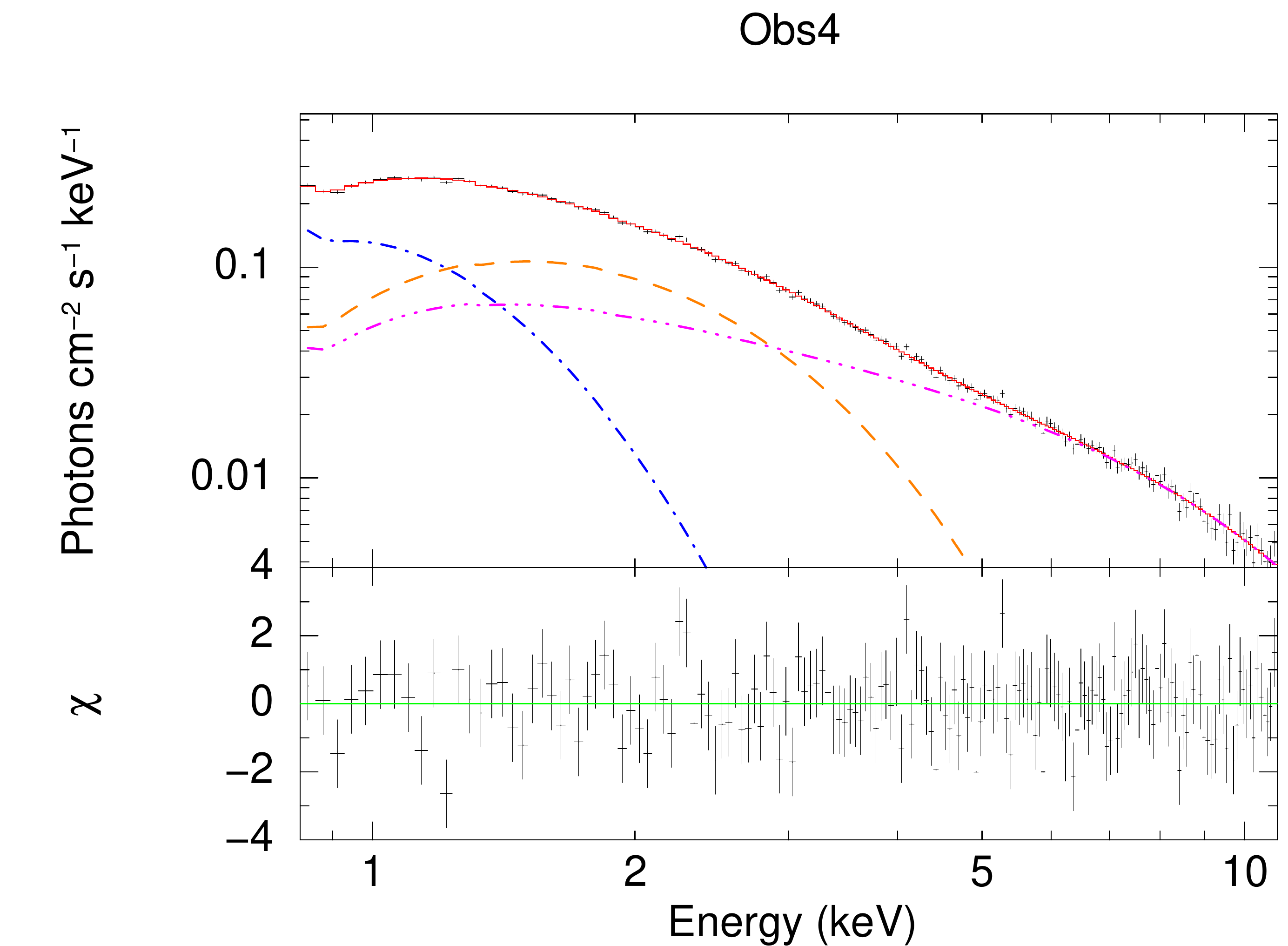}
\caption{Typical phase resolved spectra from the first phase bin of 4 observations with mHz QPO detections. The blue dash-dotted line, the orange dash line, and purple dash-dotted line represent {\sc DISKBB}, {\sc BBODYRAD}, and {\sc NTHCOMP}, respectively. The red solid line represents the sum of these three components.}
\label{fig5}
\end{figure*}

To obtain the phase-resolved spectra, we divided an oscillation cycle into 20 bins according to the phase defined by HHT, as we did for the oscillation profile, and extracted their individual energy spectra. Using XSPEC v12.10.1, these spectra were fitted with the model \citep{2015MNRAS.454..541L,2016ApJ...831...34S}:
\[
PHABS\times(BBODYRAD+DISKBB+NTHCOMP)
\]
where {\sc PHABS} is photoelectric absorption, {\sc BBODYRAD} is blackbody from the neutron-star surface, {\sc DISKBB} \citep{1984PASJ...36..741M,1986ApJ...308..635M} is multi-color disk blackbody to describe the multi-temperature thermal emissions from an accretion disc, and {\sc NTHCOMP} \citep{1996MNRAS.283..193Z,1999MNRAS.309..561Z} is inverse Compton scattering process in the corona, with the seed thermal photons coming from the accretion disk \citep{2013MNRAS.432.1144S,2014MNRAS.445.3659L}. The interstellar absorption was fixed at $0.36\times10^{22} cm^{-2}$ \citep{2013MNRAS.432.1144S}, selecting the solar abundance from \citet{2000ApJ...542..914W} and the cross-section from \citet{1996ApJ...465..487V}.  Because the mHz QPO is considered as metastable nuclear burning on neutron star surface, only the parameters in {\sc BBODYRAD} were allowed to vary for different phase bins with the parameters of other components being tied to the optimal values evaluated from the detected spectra of the whole cycle. All 20 phase bin spectra of the 4 observations are well fitted by this model, and therefore, the spectral parameter variations on the neutron star surface can be resolved for the whole complete cycle. Figure \ref{fig5} shows the typical spectra from the first phase bin of 4 observations.

Figure \ref{fig4} illustrates modulations of the two untied parameters in {\sc BBODYRAD}, the temperature $T_{bb}$ and the apparent area $R_{\infty}^{2}$, as well as oscillation profile.\ The apparent area $(R_{\infty}^{2})$ with the mHz QPO phase was calculated by the relation \citep{1985ApJ...299..487S,2016ApJ...831...34S}:
\[
R_{BB}^{2}=R_{\infty}^{2}\times f_{col}^{4}\times \left(1-2\frac{M}{R_{NS}}\right)
\]
where G=c=1.\ The source surface $(R_{BB}^{2})$ was derived from {\sc BBODYRAD} components with a distance of 6 kpc \citep{2006ApJ...639.1033G}. The factor $(f_{col})$ was set to 1.6 with a helium-enriched environment \citep{2015MNRAS.454..541L}. According to the bolometric flux oscillations that occur during the rise of X-ray burst from \emph{Rossi X-ray Timing Explorer} (\emph{RXTE}) observations, $M/R_{NS}$ is 0.126 \citep{2002ApJ...564..353N}.\  Table \ref{tab:spectrum_fit2} and Table \ref{tab:spectrum_fit} list the untied spectral papameters from {\sc BBODYRAD} and tied parameters from {\sc DISKBB} and {\sc NTHCOMP}, respectively.

Because the count rate is proportional to temperature to the power of 4 and area ($count\ rate \propto L \propto AT^{4}$). We investigated the correlations between the parameters of {\sc BBODYRAD} and the QPO-profile. Previous studies gave two contradictory results. \citet{2016ApJ...831...34S} concluded that the area variation dominates the flux oscillation of mHz QPO, whereas \citet{2018ApJ...865...63S} showed that variation is mainly owing to the temperature modulation. To further study which factor is more important to mHz QPO, the Pearson correlation coefficient, r, was applied to investigate the linear relationship between two variables (i.e. count rate vs. area and count rate vs. temperature). The corresponding p-values, under the null hypothesis that two variables follow the Gaussian distribution with zero correlation coefficient, can be used to verify if spectral parameter has strong linear relation to the oscillation count rates. The smaller p-value indicates the stronger correlation between two variables.

Table \ref{tab:correlation} lists the correlation coefficients and the corresponding p-values. For the Obs1 spectra, there is a strong correlation between $T_{bb}$ and count rates with a linear correlation coefficient r=0.84 and p-value of p=$3\times10^{-6}$ but almost no correlation between the apparent area and count rates (r=0.19, p=0.41). It indicates that the oscillation is primarily attributed to the temperature variation in an approximately constant apparent area (also refer to Figure \ref{fig4}). For the Obs2 spectra, the marginally positive correlation between $T_{bb}$ and the count rates is r=0.48 (p=0.03), and the apparent area is highly correlated with the count rates with r=0.91 (p=$2\times10^{-8}$). For the Obs3 and Obs4 spectra, in addition to the marginally positive correlations between $T_{bb}$ and the count rates, where r=0.7 (p=$5\times10^{-3}$) and r=0.66 (p=$1\times10^{-3}$) for Obs3 and Obs4, respectively, the apparent area is significantly correlated with the count rates, with r=0.82 (p=$8\times10^{-6}$) and r=0.88 (p=$2\times10^{-7}$) for Obs3 and Obs4, respectively. These results show that the area variation dominates the mHz QPO with a constant temperature in Obs 2, 3, and 4 but the temperature variation dominates the mHz QPO with constant area in Obs1. Further implications of these results will be discussed in Section \ref{sec:discussion}. 

\section{Discussion} \label{sec:discussion}

We have utilized HHT to characterize the HHT-based timing properties, extracted the 4U 1636-53 $\sim$8 mHz QPOs’ instantaneous phase, and constructed the modulation profiles and phase-resolved spectra for the complete cycle. The mHz QPO is a nonstationary periodic signal, for which the frequency or the amplitude would change with time. Because the dynamic power spectra only obtained the mHz QPO frequency in a time interval, it cannot precisely derive the instantaneous frequency. Conversely, the HHT technique enables us to determine the instantaneous frequency for each data point, that provides an alternative point of view to study the nature of the mHz QPOs. More important is that the HHT may resolve the instantaneous phase, that is believed to be closely related to the physics behind the oscillations rather than the time. With the instantaneous phase, phase-resolved analyses can be processed \citep[e.g.][]{2014ApJ...788...31H}.

In this work, we extracted the instantaneous phase of mHz QPO of 4U 1636-53. Spectral parameter variations can be resolved for the whole mHz oscillation cycle. Our phase-resolved analyses that employ the HHT in this work have a phase resolution of 0.05 cycles for a complete QPO cycle. It shows that HHT has the ability to provide more details in the variations of the spectral parameters. In theoretical model, mHz QPO attributes to the variation of nuclear reaction rate modulations in the hot spot \citep{2007ApJ...665.1311H} that may cause the temperature oscillation or apparent area variation. Recently, there are two contradictory results for mHz QPO observations. \citet{2018ApJ...865...63S} revealed significant oscillations at the frequency $\sim$8 mHz in GS 1826-238 from the \emph{Neutron Star Interior Composition Explorer} (\emph{NICER}) observation on September 9, 2017. This $\sim$8 mHz frequency and its amplitude are consistent with other accreting neutron star systems with mHz QPO detections. Their phase-resolved spectra show that the mHz oscillation is produced by modulation of the temperature component of blackbody emission from the neutron star surface with a constant emission area through the oscillation cycle. However, \citet{2016ApJ...831...34S} provided a different conclusion for the mHz QPO of 4U1636-53. Their phase-resolved spectra showed that mHz QPOs are owing to the periodically changing size of the hot spot with a constant temperature. In this work, the detailed and more precise phase resolved spectra can exhibit the correlations shown in Section 3.2. The result reveals that the area variation dominates the mHz QPO in Obs 2, 3 and 4, but temperature variation has much stronger correlation with the oscillation count rate than the apparent area in Obs1. This indicates that either temperature or apparent area variations of the hot spot may dominate the mHz QPO oscillation. The cause of the difference is not clear but it might be related to the spectral state of the source. Figure 1 in \citet{2015MNRAS.454..541L} showed the spectral states of these 4 observations in the color-color diagram. We find that the hard color of Obs1 is higher than the others. This is a possible clue why the Obs1 behaves differently than the other three. However, because there are only 4 observations with mHz QPO detections, more observations are required to verify if the different behavior of Obs1 is due to the spectral state or just a coincident.

One of the important factors in marginally stable burning is the accretion rate. The theoretical interpretation showed that the accretion rate is close to the Eddington accretion rate ($\sim\dot M_{Edd}$) \citep{2007ApJ...665.1311H}, but current observations indicated that the accretion rate is much lower ($\sim0.1 \dot M_{Edd}$) for the sources with mHz QPOs. \citet{1981ApJ...247..267F} suggested that the burst ignition of neutron star depends on the accretion rate per unit area, $\dot m$ instead of the overall accretion rate. The local accretion rate ($\dot m$) does not need to be the same everywhere on the neutron star surface \citep{1998ASIC..515..419B}. The nuclear burning depends on the local accretion rate \citep{1998ASIC..515..419B,2007ApJ...665.1311H}. For a fast-spinning neutron star, a latitude change in the local accretion rate is influenced by the variation in the effective surface gravity from the equator to the pole \citep{2018ApJ...865...63S}. \citet{2016MNRAS.463.2358L} discovered that mHz QPOs only associate with bursts with positive convexity. Because \citet{2008MNRAS.383..387M} found that type-I X-ray bursts that ignite at the equator always have positive convexity, the mHz QPO locates on the equator of the neutron star. \citet{2018ApJ...865...63S} roughly estimate the effective surface gravity of the neutron star in 4U1636-53; the value on the pole can be $\sim11\%$ stronger than the one on the equator. They found this value is sufficient for influencing marginally stable burning by effecting the local accretion rate, and marginally stable burning on the equator belt on the neutron star surface is possible \citep{2008ApJ...673L..35A,2018ApJ...865...63S}. Even though using the local accretion rate can explain marginally stable nuclear burning, a more complete theoretical model is required to explain the phenomena that we observed. \citet{2007ApJ...665.1311H} calculated that marginally stable burning is one-dimensional; thus, the burning area is not explored \citep{2018ApJ...865...63S}.

Data collected by \emph{XMM-Newton} were analyzed in this work. The mHz QPOs are typically detected in low energy bands (1-5 keV); thus, the energy range of \emph{XMM-Newton} (0.1 to 15 keV) is beneficial for observing the mHz QPOs. \emph{XMM-Newton} is a high altitude satellite with an orbital period of 2789.6 minutes, that enables it to observe a source for a long time without interruption. Because HHT prefers continuous data, the data collected by \emph{XMM-Newton} are beneficial for using the HHT to analyze mHz QPOs. The high spectrum resolution of \emph{XMM-Newton} is also advantageous to the spectral analysis. In addition to \emph{XMM-Newton}, \emph{Nuclear Spectroscopic Telescope Array mission} (\emph{NuSTAR}) and \emph{NICER} are possible choices that can be used to study mHz QPOs. Unfortunately, because \emph{NuSTAR} focuses on high X-ray energy bands (3-79 keV), it may not be suitable for studying mHz QPOs (typically detected in lower energy bands). The time resolution of \emph{NICER} is 100 nsec, and its spectral band overlaps \emph{RXTE} and \emph{XMM-Newton} (0.2-12 keV). The high time resolution enables us to analyze more timing properties by the HHT. \emph{NICER} has a large effective area, that is beneficial for collecting photons. More photons have more timing information; thus, \emph{NICER} would be a reasonable choice to perform timing analysis. However, \emph{NICER} is a facility on-board the International Space Station (ISS), and the observations are usually interrupted due to Earth occultation for each ISS orbital cycle ($\sim90$ min). On the other hand, X-ray telescopes that are expected to be operated in the next decade, such as \emph{enhanced X-ray Timing and Polarimetry mission} (\emph{eXTP}) and \emph{X-ray Imaging and Spectroscopy Mission} (\emph{XRISM}), although they have better time and spectral resolution in the low energy band ($\sim0.5-10$ keV), the Earth occultation would induce too much alias in the HHT analysis for these low-altitude satellites, similar to \emph{NICER}. These make the data collected by \emph{XMM-Newton} the best data for analysis by the HHT for the mHz QPO investigations. In the future, more \emph{XMM-Newton} observations will enable us to investigate the physical reasons of mHz QPOs by using the HHT to analyze more timing properties and spectral modulation of mHz QPOs.

%\url{https://github.com/AASJournals/Tutorials/tree/master/Repositories}.

\acknowledgments
We would like to thank anonymous reviewer for helpful and valuable comments on the manuscript. We would like to thank Dr.\ Yi-Hao Su for useful advice regarding the HHT analysis.\ This work is based on observations obtained with \emph{XMM-Newton}, an ESA science mission with instruments and contributions directly funded by ESA Member States and the USA (NASA).\ This work has made use of data obtained from the High Energy Astrophysics Science Archive Research Center (HEASARC), provided by NASA's Goddard Space Flight Center.\ The HHT codes were provided by the Research Center for Adaptive Data Analysis in the National Central University of Taiwan.\ This research was supported by the  grants MOST 108-2112-M-008-005- and 109-2112-M-008-004- from the Ministry of Science and Technology of Taiwan.
%We thank all the people that have made this AASTeX what it is today.  This
%includes but not limited to Bob Hanisch, Chris Biemesderfer, Lee Brotzman,
%Pierre Landau, Arthur Ogawa, Maxim Markevitch, Alexey Vikhlinin and Amy
%Hendrickson. Also special thanks to David Hogg and Daniel Foreman-Mackey
%for the new "modern" style design. Considerable help was provided via bug
%reports and hacks from numerous people including Patricio Cubillos, Alex
%Drlica-Wagner, Sean Lake, Michele Bannister, Peter Williams, and Jonathan
%Gagne.


\vspace{5mm}
\facilities{ADS, HEASARC, \emph{XMM-Newton}}


\software{Science Analysis System (SAS) v.16.0.0. \citep{2017xru..conf...84G},
			   XSPEC \citep{1996ASPC..101...17A},
			   astropy \citep{2013A&A...558A..33A}
          }




\bibliography{my_citation}{}
\bibliographystyle{aasjournal}


\begin{table}[h!]
\centering
\caption{\emph{XMM-Newton} observations of 4U1636-53 from 2000 to 2015.} % needs to go inside longtable environment
\begin{tabular}[c]{c c c c c} 
\hline
Observation ID & Observation date & PN exposure time & Time span of QPO & Note\\
{} & {} & (seconds) & (seconds) & {} \\ \hline
0105470101 & 2000-09-07 & 23607 & No detection & -\\
\hline
0105470401 & 2001-08-24 & 21601 & No detection & -\\
\hline
0303250201 & 2005-08-29 & 31336 & No detection & -\\
\hline
0500350301 & 2007-09-28 & 31935 & 5420 & Obs1 \\
\hline
0500350401 & 2008-02-27 & 39942 & 16340 & Obs2 \\
\hline
0606070101 & 2009-03-14 & 41179 & 13317 & Obs3 \\
\hline
0606070201 & 2009-03-25 & 28939 & No detection & -\\
\hline
0606070301 & 2009-09-05 & 43200 & 11949 & Obs4 \\
\hline
0606070401 & 2009-09-11 & 27239 & No detection & -\\	
\hline
0764180201 & 2015-08-25 & 40800 & No detection & -\\
\hline
0764180301 & 2015-09-05 & 36899 & No detection & -\\	
\hline
0764180401 & 2015-09-18 & 37100 & No detection & -\\ \hline
\label{tab:ID}
\end{tabular}
\end{table}

\begin{table}[h!]
\centering
\caption{All {\sc BBODYRAD} parameters in this work.} % needs to go inside longtable environment
\begin{tabular}{@{\extracolsep{\fill}}c c c c c || c c c c c@{}} 

\hline
Note & Phase & Temperature (keV) & Area normalized & $\chi^{2}/d.o.f$ & Note & Phase & Temperature (keV) & Area normalized & $\chi^{2}/d.o.f$\\
\hline
Obs1 & 1 & $0.5477_{-0.0080}^{+0.0080}$ & $789_{-34}^{+35}$ & $138.46/155$ & Obs2 & 1 & $0.5952_{-0.0053}^{+0.0038}$ & $665_{-22}^{+19}$ & $173.10/169$\\
\hline
Obs1 & 2 & $0.5507_{-0.0083}^{+0.0083}$ & $757_{-34}^{+34}$ & $155.41/159$ & Obs2 & 2 & $0.5971_{-0.0052}^{+0.0038}$ & $639_{-21}^{+19}$ & $165.19/170$\\
\hline
Obs1 & 3 & $0.5472_{-0.0084}^{+0.0085}$ & $764_{-34}^{+35}$ & $128.89/156$ & Obs2 & 3 & $0.5913_{-0.0055}^{+0.0039}$ & $644_{-22}^{+19}$ & $185.82/171$\\
\hline
Obs1 & 4 & $0.5430_{-0.0090}^{+0.0090}$ & $767_{-18}^{+36}$ & $155.12/156$ & Obs2 & 4 & $0.5895_{-0.0056}^{+0.0040}$ & $647_{-22}^{+19}$ & $154.66/171$\\
\hline
Obs1 & 5 & $0.5412_{-0.0091}^{+0.0091}$ & $756_{-29}^{+36}$ & $172.28/157$ & Obs2 & 5 & $0.5909_{-0.0057}^{+0.0040}$ & $625_{-22}^{+19}$ & $210.04/170$\\
\hline
Obs1 & 6 & $0.5390_{-0.0092}^{+0.0092}$ & $761_{-35}^{+36}$ & $119.45/159$ & Obs2 & 6 & $0.5948_{-0.0052}^{+0.0040}$ & $606_{-25}^{+19}$ & $199.24/170$\\
\hline
Obs1 & 7 & $0.5377_{-0.0095}^{+0.0094}$ & $762_{-36}^{+36}$ & $149.83/159$ & Obs2 & 7 & $0.5904_{-0.0057}^{+0.0040}$ & $618_{-26}^{+19}$ & $160.11/171$\\
\hline
Obs1 & 8 & $0.5387_{-0.0093}^{+0.0092}$ & $748_{-35}^{+35}$ & $148.07/157$ & Obs2 & 8 & $0.5917_{-0.0058}^{+0.0040}$ & $604_{-26}^{+19}$ & $121.97/171$\\
\hline
Obs1 & 9 & $0.5381_{-0.0095}^{+0.0095}$ & $754_{-36}^{+36}$ & $134.25/159$ & Obs2 & 9 & $0.5960_{-0.0056}^{+0.0040}$ & $594_{-21}^{+19}$ & $166.33/171$\\
\hline
Obs1 & 10 & $0.5468_{-0.0091}^{+0.0091}$ & $709_{-33}^{+34}$ & $157.38/158$ & Obs2 & 10 & $0.5932_{-0.0058}^{+0.0040}$ & $599_{-19}^{+19}$ & $166.44/177$\\
\hline
Obs1 & 11 & $0.5458_{-0.0089}^{+0.0089}$ & $721_{-34}^{+34}$ & $178.75/157$ & Obs2 & 11 & $0.5893_{-0.0059}^{+0.0041}$ & $606_{-21}^{+19}$ & $157.21/172$\\
\hline
Obs1 & 12 & $0.5418_{-0.0093}^{+0.0093}$ & $737_{-18}^{+35}$ & $143.77/158$ & Obs2 & 12 & $0.5896_{-0.0058}^{+0.0040}$ & $606_{-26}^{+19}$ & $185.17/172$\\
\hline
Obs1 & 13 & $0.5323_{-0.0099}^{+0.0099}$ & $774_{-37}^{+38}$ & $141.56/157$ & Obs2 & 13 & $0.5897_{-0.0058}^{+0.0040}$ & $612_{-21}^{+19}$ & $154.24/172$\\
\hline
Obs1 & 14 & $0.5312_{-0.0101}^{+0.0100}$ & $793_{-37}^{+39}$ & $170.59/157$ & Obs2 & 14 & $0.5903_{-0.0059}^{+0.0041}$ & $607_{-21}^{+19}$ & $161.58/171$\\
\hline
Obs1 & 15 & $0.5363_{-0.0093}^{+0.0093}$ & $769_{-36}^{+36}$ & $138.23/156$ & Obs2 & 15 & $0.5965_{-0.0051}^{+0.0039}$ & $592_{-26}^{+19}$ & $178.58/172$\\
\hline
Obs1 & 16 & $0.5420_{-0.0091}^{+0.0090}$ & $741_{-32}^{+35}$ & $135.26/158$ & Obs2 & 16 & $0.5967_{-0.0056}^{+0.0040}$ & $596_{-26}^{+19}$ & $181.53/170$\\
\hline
Obs1 & 17 & $0.5538_{-0.0082}^{+0.0082}$ & $731_{-33}^{+30}$ & $154.51/157$ & Obs2 & 17 & $0.5937_{-0.0050}^{+0.0039}$ & $633_{-27}^{+19}$ & $196.67/170$\\
\hline
Obs1 & 18 & $0.5516_{-0.0081}^{+0.0082}$ & $765_{-33}^{+34}$ & $171.72/156$ & Obs2 & 18 & $0.5914_{-0.0053}^{+0.0041}$ & $662_{-22}^{+19}$ & $184.78/171$\\
\hline
Obs1 & 19 & $0.5565_{-0.0081}^{+0.0080}$ & $748_{-33}^{+30}$ & $149.87/157$ & Obs2 & 19 & $0.5948_{-0.0048}^{+0.0037}$ & $658_{-25}^{+19}$ & $192.30/171$\\
\hline
Obs1 & 20 & $0.5591_{-0.0075}^{+0.0076}$ & $747_{-32}^{+32}$ & $160.29/156$ & Obs2 & 20 & $0.5975_{-0.0050}^{+0.0036}$ & $661_{-22}^{+19}$ & $202.71/171$\\
\hline
\hline
Obs3 & 1 & $0.5398_{-0.0047}^{+0.0048}$ & $714_{-30}^{+30}$ & $180.53/171$ & Obs4 & 1 & $0.5637_{-0.0056}^{+0.0058}$ & $820_{-20}^{+23}$ & $154.50/169$\\
\hline
Obs3 & 2 & $0.5388_{-0.0048}^{+0.0049}$ & $697_{-31}^{+30}$ & $189.44/170$ & Obs4 & 2 & $0.5591_{-0.0059}^{+0.0061}$ & $818_{-23}^{+23}$ & $190.83/169$\\
\hline
Obs3 & 3 & $0.5270_{-0.0051}^{+0.0052}$ & $737_{-31}^{+31}$ & $224.64/170$ & Obs4 & 3 & $0.5549_{-0.0062}^{+0.0063}$ & $825_{-24}^{+24}$ & $177.72/169$\\
\hline
Obs3 & 4 & $0.5240_{-0.0053}^{+0.0055}$ & $737_{-31}^{+31}$ & $188.45/171$ & Obs4 & 4 & $0.5597_{-0.0062}^{+0.0063}$ & $776_{-23}^{+23}$ & $184.26/170$\\
\hline
Obs3 & 5 & $0.5292_{-0.0052}^{+0.0054}$ & $701_{-31}^{+30}$ & $155.28/171$ & Obs4 & 5 & $0.5584_{-0.0064}^{+0.0066}$ & $767_{-23}^{+23}$ & $156.55/169$\\
\hline
Obs3 & 6 & $0.5289_{-0.0055}^{+0.0056}$ & $676_{-31}^{+30}$ & $180.79/170$ & Obs4 & 6 & $0.5530_{-0.0033}^{+0.0067}$ & $796_{-23}^{+24}$ & $193.24/170$\\
\hline
Obs3 & 7 & $0.5291_{-0.0055}^{+0.0057}$ & $668_{-31}^{+30}$ & $150.00/170$ & Obs4 & 7 & $0.5549_{-0.0064}^{+0.0066}$ & $777_{-23}^{+23}$ & $201.52/170$\\
\hline
Obs3 & 8 & $0.5293_{-0.0055}^{+0.0057}$ & $659_{-31}^{+30}$ & $199.10/170$ & Obs4 & 8 & $0.5572_{-0.0032}^{+0.0066}$ & $767_{-23}^{+23}$ & $123.72/170$\\
\hline
Obs3 & 9 & $0.5288_{-0.0054}^{+0.0056}$ & $677_{-31}^{+30}$ & $189.07/171$ & Obs4 & 9 & $0.5615_{-0.0064}^{+0.0066}$ & $734_{-22}^{+23}$ & $161.16/170$ \\
\hline
Obs3 & 10 & $0.5317_{-0.0055}^{+0.0057}$ & $652_{-30}^{+30}$ & $146.95/171$ & Obs4 & 10 & $0.5559_{-0.0064}^{+0.0067}$ & $763_{-23}^{+23}$ & $153.28/168$\\
\hline
Obs3 & 11 & $0.5287_{-0.0058}^{+0.0059}$ & $661_{-31}^{+30}$ & $180.95/171$ & Obs4 & 11 & $0.5595_{-0.0064}^{+0.0066}$ & $740_{-23}^{+23}$ & $170.36/170$\\
\hline
Obs3 & 12 & $0.5271_{-0.0055}^{+0.0056}$ & $672_{-31}^{+30}$ & $176.76/170$ & Obs4 & 12 & $0.5549_{-0.0067}^{+0.0069}$ & $754_{-23}^{+24}$ & $207.85/168$\\
\hline
Obs3 & 13 & $0.5287_{-0.0056}^{+0.0058}$ & $660_{-31}^{+30}$ & $189.05/171$ & Obs4 & 13 & $0.5605_{-0.0064}^{+0.0065}$ & $737_{-23}^{+23}$ & $165.73/170$\\
\hline
Obs3 & 14 & $0.5268_{-0.0056}^{+0.0058}$ & $683_{-31}^{+31}$ & $208.90/172$ & Obs4 & 14 & $0.5535_{-0.0066}^{+0.0068}$ & $767_{-23}^{+24}$ & $174.87/172$\\
\hline
Obs3 & 15 & $0.5368_{-0.0055}^{+0.0056}$ & $635_{-31}^{+30}$ & $179.65/171$ & Obs4 & 15 & $0.5587_{-0.0064}^{+0.0066}$ & $755_{-23}^{+23}$ & $168.19/168$\\
\hline
Obs3 & 16 & $0.5322_{-0.0054}^{+0.0055}$ & $670_{-31}^{+31}$ & $141.18/172$ & Obs4 & 16 & $0.5585_{-0.0063}^{+0.0065}$ & $778_{-12}^{+23}$ & $145.32/169$\\
\hline
Obs3 & 17 & $0.5322_{-0.0052}^{+0.0053}$ & $697_{-31}^{+31}$ & $202.26/172$ & Obs4 & 17 & $0.5606_{-0.0060}^{+0.0061}$ & $792_{-23}^{+24}$ & $200.99/168$\\
\hline
Obs3 & 18 & $0.5351_{-0.0049}^{+0.0050}$ & $719_{-31}^{+31}$ & $170.80/171$ & Obs4 & 18 & $0.5631_{-0.0057}^{+0.0059}$ & $800_{-20}^{+12}$ & $163.63/170$\\
\hline
Obs3 & 19 & $0.5407_{-0.0047}^{+0.0048}$ & $723_{-31}^{+30}$ & $173.93/171$ & Obs4 & 19 & $0.5624_{-0.0056}^{+0.0057}$ & $831_{-23}^{+23}$ & $203.28/169$\\
\hline
Obs3 & 20 & $0.5360_{-0.0047}^{+0.0048}$ & $752_{-31}^{+31}$ & $173.03/171$ & Obs4 & 20 & $0.5659_{-0.0056}^{+0.0058}$ & $814_{-23}^{+21}$ & $158.42/170$\\
\hline
\end{tabular}
\label{tab:spectrum_fit2}
\end{table}

\begin{table}[h!]
\centering
\caption{Spectral parameters of DISKBB and NTHCOMP.} % needs to go inside longtable environment
\begin{tabular}[c]{ c c c c c c } 
\hline
Component & Parameter & Obs1 & Obs2 & Obs3 & Obs4 \\ \hline
{\sc DISKBB} & $\kappa T_{0}$ (keV) & $0.277_{-0.007}^{+0.007}$ & $0.275_{-0.005}^{+0.005}$ & $0.233_{-0.007}^{+0.007}$ & $0.287_{-0.006}^{+0.006}$ \\
{        } & $Nor$ & $9441_{-738}^{+842}$ & $10125_{-526}^{+576}$ & $16454_{-1550}^{+1765}$ & $8126_{-514}^{+551}$ \\ 
\hline
{\sc NTHCOMP} & $NTHCOMP (\Gamma)$ & $1.483_{-0.057}^{+0.054}$ & $1.549_{-0.026}^{+0.025}$ & $1.727_{-0.024}^{+0.021}$ & $1.484_{-0.039}^{+0.036}$ \\
{         } & $kT_{e} (keV)$ & $2.196_{-0.071}^{+0.076}$ & $2.334_{-0.037}^{+0.039}$ & $2.751_{-0.071}^{+0.073}$ & $2.215_{-0.047}^{+0.049}$ \\
{         } & $Nor$ & $0.113_{-0.019}^{+0.020}$ & $0.198_{-0.016}^{+0.017}$ & $0.318_{-0.023}^{+0.023}$ & $0.134_{-0.016}^{+0.017}$ \\
\hline
$\chi^{2}/d.o.f$ & { } & 3100.6/3144 & 3874.6/3638 & 3729.8/3417 & 3546.7/3387 \\ \hline
\end{tabular}
\tablenotetext{a}{$n_{H}$ is fixed at $0.36 \times 10^{22} cm^{-2}.$}
\label{tab:spectrum_fit}
\end{table}

\begin{table}[h!]
\centering
\caption{The correlation between count rate and temperature (T) and between count rate and area (A).} % needs to go inside longtable environment
\begin{tabular}[c]{| c | c | c | c | c |} 
\hline
 Note & T:r-value & T:p-value & A:r-value & A:p-value\\ 
 \hline
 Obs1 & 0.84 & $3\times10^{-6}$ & 0.19 & 0.41 \\
 \hline
 Obs2 & 0.48 & 0.03 & 0.91 & $2\times10^{-8}$ \\
 \hline
 Obs3 & 0.70 & 0.005 & 0.82 & $8\times10^{-6}$ \\
 \hline
 Obs4 & 0.66 & 0.001 & 0.88 & $2\times10^{-7}$ \\
 \hline
 \end{tabular}
\label{tab:correlation}
\end{table}

%orign table
%\setlength\LTleft{0pt}
%\setlength\LTright{0pt}
%\begin{longtable}[!hbp]{@{\extracolsep{\fill}}c c c c c | c c c c c@{}}

%\caption{All {\sc BBODYRAD} components in this work.} % needs to go inside longtable environment
%\label{tab:spectrum_fit2}
%\end{longtable}

%\begin{deluxetable*}{cccccccc}
%\tablenum{3}
%\tablecaption{The fitting result of 4U1636-53. $n_{H}$ is fixed on $0.36 \times 10^{22} cm^{-2}.$\label{tab:spectrum_fit2}}
%\tablewidth{0pt}
%\tablehead{
%\colhead{Note} & \colhead{Phase} & \colhead{Temperature (keV)} & \colhead{Area normalized} \\
%}
%\startdata

%\enddata
%\end{deluxetable*}

\end{document}